\documentclass[a4paper,USenglish,cleveref, autoref, thm-restate,runningheads]{llncs}

\usepackage[dvipsnames]{xcolor}
\usepackage{todonotes}
\usepackage{tikz}
\usetikzlibrary{arrows,shapes,snakes,automata,backgrounds,petri}
\usetikzlibrary{shapes.multipart}
\usepackage{comment}
\usepackage[noend]{algpseudocode}
\usepackage{caption}
\usepackage{subcaption}

\usepackage[utf8]{inputenc}          
\usepackage[T1]{fontenc}             
\usepackage[USenglish]{babel}        
\usepackage[intlimits]{amsmath}      
\usepackage{amsfonts}                
\usepackage{amssymb}                 
\usepackage{amsmath}                 

\usepackage{mathtools}
\usepackage{setspace}
\usepackage{graphicx}
\usepackage{hyperref}

\usepackage{algorithm}
\usepackage[noend]{algpseudocode}

\newcommand{\DS}{\textsc{Dominating Set}}
\newcommand{\EDS}{\textsc{Extension Dominating Set}}

\newcommand{\RD}{\textsc{Roman Domination}}

\newcommand{\NP}{\textsf{NP}}

\newcommand{\RomanUpperbound}{1.9332}

\newcommand{\Oh}{\mathcal{O}}

\newenvironment{pf}{\begin{proof}}{\hfill\qed\end{proof}}

\newtheorem{observation}[theorem]{Observation}

\newtheorem{brarule}{Branching Rule}%[section]

\newtheorem{redrule}{Reduction Rule}

\newcommand{\Active}{A}

\usepackage[utf8]{inputenc}          % um Umlaute direkt eingeben zu können
\usepackage[T1]{fontenc}             % für "modernere" Schriftkodierung

\usepackage{mathtools}
\usepackage{setspace}
\usepackage{graphicx}
\usepackage{hyperref}

\newcommand{\RomanChordalUpperbound}{1.8940}
\newcommand{\RomanIntervalUpperbound}{1.7321}\newcommand{\RomanIntervalExact}{\sqrt{3}}
\newcommand{\RomanSplitUpperbound}{1.4656}

\usepackage{xspace}

\usepackage[nocompress]{cite}

\makeatletter
\def\BState{\State\hskip-\ALG@thistlm}
\makeatother

\begin{document}

\title{Roman Census: Enumerating and Counting Roman Dominating Functions on Graph Classes}

\titlerunning{Roman Census}
\author{Faisal N. Abu-Khzam\inst1\orcidID{0000-0001-5221-8421} 
\and Henning Fernau\inst2\orcidID{0000-0002-4444-3220}
\and Kevin Mann\inst2\orcidID{0000-0002-0880-2513} 
}
\authorrunning{F. Abu-Khzam, H. Fernau, and K. Mann}
\institute{
Department of Computer Science and Mathematics\\
Lebanese American University, 
Beirut, Lebanon.\\
\email{faisal.abukhzam@lau.edu.lb}\\
\and
Universit\"at Trier, Fachbereich~4 -- Abteilung Informatikwissenschaften\\  
54286 Trier, Germany.\\
\email{\{fernau,mann\}@uni-trier.de}
}
\maketitle 

%\keywords{special graph classes, counting problems, enumeration problems, domination problems, Roman domination}

\begin{abstract}
The concept of Roman domination has recently been studied concerning enumerating and counting (WG 2022). It has been shown that minimal Roman dominating functions can be enumerated with polynomial delay, contrasting what is known about minimal dominating sets. The running time of the algorithm could be estimated as $\Oh(\RomanUpperbound^n)$ on general graphs of order~$n$.
In this paper, we focus on special graph classes. More specifically,  for chordal graphs, we present an enumeration algorithm running in time $\Oh(\RomanChordalUpperbound^n)$. For interval graphs, we can lower this time further to $\Oh(\RomanIntervalUpperbound^n)$. Interestingly, this also matches (exactly) the known lower bound. We can also provide a matching lower and upper bound for forests, which is (incidentally) the same, namely $\Oh(\RomanIntervalExact^n)$.
Furthermore,
we show an enumeration algorithm running in time $\Oh(\RomanSplitUpperbound^n)$
for split graphs and for cobipartite graphs.
Our approach also allows to give concrete formulas for counting minimal Roman dominating functions
on special graph families like paths.
\end{abstract}

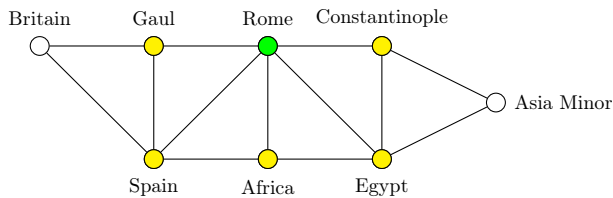
\begin{figure}[bth]
\begin{center}
	\begin{tikzpicture}[transform shape,scale=0.75]
			\tikzset{every node/.style={circle,minimum size=0.3cm}}
			\node[draw] (Britain) at (-4,1) {};
			{\node[draw] (Gaul) at (-2,1) {};
			\node[draw] (Spain) at (-2,-1) {};
			\node[draw] (Rome) at (0,1) {};
			\node[draw] (Africa) at (0,-1) {};
			\node[draw] (Constantinople) at (2,1) {};
			\node[draw] (Egypt) at (2,-1) {};}
			{\node[draw,fill=yellow] at (-2,1) {};
			\node[draw,fill=yellow] at (-2,-1) {};
			\node[draw,fill=green] at (0,1) {};
			\node[draw,fill=yellow] at (0,-1) {};
			\node[draw,fill=yellow] at (2,1) {};
			\node[draw,fill=yellow] at (2,-1) {};}
			\node[draw,label={right:Asia Minor}] (Asia) at (4,0) {};
			\node at (2,1.5) {Constantinople};
			\node at (-4,1.5) {Britain};
			\node at (-2,1.5) {Gaul};
			\node at (0,1.5) {Rome};
			\node at (-2,-1.5) {Spain};
			\node at (0,-1.5) {Africa};
			\node at (2,-1.5) {Egypt};
			\path (Britain) edge[-] (Gaul);
			\path (Britain) edge[-] (Spain);
			\path (Gaul) edge[-] (Spain);
			\path (Rome) edge[-] (Spain);
			\path (Rome) edge[-] (Gaul);
			\path (Rome) edge[-] (Africa);
			\path (Rome) edge[-] (Constantinople);
			\path (Rome) edge[-] (Egypt);
			\path (Africa) edge[-] (Spain);
			\path (Africa) edge[-] (Egypt);
			\path (Asia) edge[-] (Egypt);
			\path (Asia) edge[-] (Constantinople);
			\path (Egypt) edge[-] (Constantinople);
        \end{tikzpicture}
%\includegraphics[width=.65\textwidth]{Roman-mymap.pdf}
%\epsfbox{Roman-mymap.eps}
\end{center}
\caption{\label{fig-Roman-map}The Roman Empire in the times of Constantine: Putting 2 armies on Rome also secures all neighboring regions (colored yellow), leaving only two regions to be secured by one army each.}
\end{figure}

\section{Introduction}

\RD\ comes with a nice (hi)story, on how to position armies on the various regions to secure the Roman Empire. `To secure' means that either
(1) a region $r$ has at least one army or 
(2) a region~$r'$  neighboring $r$
contains two armies, so that it can afford sending one army to the region $r$ without diminishing $r'$'s self-defense capabilities.
More specifically, Emperor 
Constantine had a look at a map of his empire (as discussed in~\cite{Ste99}, also see \autoref{fig-Roman-map}).\footnote{
The historical background is also nicely described
in the online Johns Hopkins Magazine,  visit 
\url{http://www.jhu.edu/~jhumag/0497web/locate3.html} to pre-view~\cite{ReVRos2000}.} 

\RD\ has received notable attention during the last two decades~\cite{Ben2004,ChaCCKLP2013,Dre2000a,Fer08,Lie2007,Lieetal2008,LiuCha2013,Pagetal2002,PenTsa2007,ShaWanHu2010}.

Relevant to our work is the development of exact algorithms. The algorithm presented in~\cite{ShiKoh2014} combines ideas from \cite{Lie2007,Roo2011} and solves the (optimization) problem in $\mathcal{O}(1.5014^n)$ time (and space).  
More combinatorial studies can be found in \cite{Chaetal2009,CheHHHM2016,Favetal2009,HedRSW2013,KraPavTap2012,LiuCha2012,LiuCha2012a,MobShe2008,XinCheChe2006,XueYuaBao2009,YerRod2013a} as well as the more recent chapter on Roman domination in~\cite{HayHedHen2020}. 
Another interesting related notion is the \emph{differential} of a graph, introduced in~\cite{Masetal2006}. See also \cite{BerFerSig2014}, for further algorithmic thoughts, as noted in \cite{AbuBCF2016,BerFer2014,BerFer2015}. 
To briefly summarize all these findings, one can say that in many ways concerning complexity, \RD\ and \DS\ behave exactly the same way. There are two notable and related exceptions, as delineated in~\cite{AbuFerMan2022}, concerning \emph{extension problems} and \emph{output-sensitive enumeration}.

\emph{Extension problems} often arise from search-tree algorithms for their optimization counterpart as follows. Assume that a search-tree node corresponds to a partial solution (or pre-solution) $U$ and instead of proceeding with the search-tree algorithm (by exploring all the possible paths from this node onward) we ask whether we can extend $U$ to a meaningful solution~$S$. In the case of \DS, this means that $S$ is an inclusion-wise minimal dominating set that contains $U$.
Unfortuntately, this \EDS\ problem and many similar problems are \NP-hard, see \cite{Bazetal2018,BonDHR2019,CasFKMS2019a,CasFGMS2021,KanLMNU2015,KanLMNU1516,Mar2013a}.
Even worse: when parameterized by the `pre-solution size,' \EDS\ is one of the few problems known to be complete for the parameterized complexity class \textsf{W}[3], as shown in~\cite{BlaFLMS2022}. 
This blocks any progress on the \textsc{Hitting Set Transversal Problem} by using extension test algorithms, which is the question whether all minimal hitting sets of a hypergraph can be enumerated with polynomial delay (or even output-polynomial) only. This question is open for four decades by now and is equivalent to several enumeration problems in logic, database theory and also to enumerating minimal dominating sets in graphs, see \cite{CreKPSV2019,EitGot95,GaiVer2017,KanLMN2014}.

By way of contrast and quite surprisingly, 
with an appropriate definition of the notion of minimality, the extension variant of  \RD\ is solvable in polynomial-time as shown in \cite{AbuFerMan2022}. This enabled us to show that enumerating all minimal Roman dominating functions is possible with polynomial delay.
This also triggered our interest in looking further into enumerating minimal Roman dominating functions on graph classes, as also done in the case of \DS, see \cite{AbuHeg2016,CouHHK2013,CouLetLie2015,GolHKKV2016,GolHegKra2016}. The basis of the output-sensitive enumeration result of~\cite{AbuFerMan2022} was several combinatorial observations. 
Here, we find ways how to use these underlying combinatorial ideas for non-trivial enumeration algorithms for minimal Roman dominating functions in split graphs, cobipartite graphs, interval graphs, forests and chordal graphs and for counting these exactly for paths. All these graph classes will be explained in separate sections below. These  exploits constitute the  main results of this paper. More details can be found at the end of the next section.

\section{Definitions and Known Results}

Let $\mathbb{N}=\{1,2,3,\dots\}$ be the set of positive integers. For $n\in\mathbb{N}$, let $[n]=\{m\in\mathbb{N}\mid m\leq n\}$.
We only consider undirected simple graphs. 
Let $G=\left(V,E\right)$ be a graph. For $U\subseteq V$, $G[U]$ denotes the graph induced by~$U$. 
For $v\in V$, $N_G(v)\coloneqq\{u\in V\mid \lbrace u,v\rbrace\in E\}$ denotes the \emph{open neighborhood} of~$v$, while $N_G[v]\coloneqq N_G(v)\cup\{v\}$ is the \emph{closed neighborhood}  of~$v$. We extend such set-valued functions $X:V\to 2^V$ to $X:2^V\to 2^V$ by setting $X(U)=\bigcup_{u\in U}X(u)$. Subset $D\subseteq V$ is a  \emph{dominating set}, or ds for short, if $N_G[D]=V$. 
For $D\subseteq V$ and $v\in D$, define the \emph{private neighborhood} of $v\in V$ with respect to~$D$ as $P_{G,D}\left( v\right)\coloneqq N_G\left[ v\right] \setminus N_G\left[D\setminus \lbrace v\rbrace\right] $.
%\\[-4.5ex]
%$$P_{G,D}\left( v\right)\coloneqq N_G\left[ v\right] \setminus \left( \bigcup_{u\in D\setminus \lbrace v\rbrace} N_G\left[ u\right]\right)\,.$$
A function $f\colon V \rightarrow \lbrace 0,1,2 \rbrace$ is called a \emph{Roman dominating function}, or rdf for short, if for each $v\in V$ with $f\left(v\right) = 0$, there exists a $u\in N_G\left( v \right)$ with $f\left(u\right)=2$. 
To simplify the notation, we define $V_i\left(f\right)\coloneqq \lbrace v\in V\mid f\left( v\right)=i\rbrace$ for $i\in\lbrace0,1,2\rbrace$. The \emph{weight} $w_f$ of a function $f\colon V \rightarrow \lbrace 0,1,2 \rbrace$ equals $|V_1|+2|V_2|$. The classical \RD\ problem asks, given $G$ and an integer $k$, if there exists an rdf  of weight at most~$k$. Connecting to the original motivation, $G$ models a map of regions, and if the region vertex~$v$ belongs to~$V_i$, then we place $i$ armies on~$v$.

%$$V_i\left(f\right)\coloneqq \lbrace v\in V\vert f\left( v\right)=i\rbrace.$$
For the definition of the problem \textsc{Extension Roman Domination}, we need to define the order $\leq$ on $\lbrace 0,1,2\rbrace^{V}$ first:
for $f,g \in \lbrace 0,1,2\rbrace^{V}$, let $f\leq g$ if and only if $f\left(v\right)\leq g\left(v\right)$ for all $v\in V$. In other words, we extend the usual linear ordering $\leq$ on $\{0,1,2\}$ to functions mapping to $\{0,1,2\}$ in a pointwise manner. 
We call a function $f\in \lbrace 0,1,2\rbrace^{V}$ a \emph{minimal Roman dominating function} if and only if $f$ is a rdf and there exists no rdf $g$, $g\neq f$, with $g\leq f$.\footnote{According to \cite{HayHedHen2020}, this notion of minimality for rdf was coined by Cockayne but then dismissed, as it does not give a proper notion of \emph{upper Roman domination} number. However, in our context, this definition seems to be most natural one; it also fits the extension framework proposed in~\cite{CasFGMS2022}.} The weights of minimal rdf can vary considerably. Consider for example a star $K_{1,n}$ with center~$c$. Then, $f_1(c)=2$, $f_1(v)=0$ otherwise; $f_2(v)=1$ for all vertices~$v$; $f_3(c)=0$, $f_3(u)=2$ for one $u\neq c$, $f_3(v)=1$ otherwise, define three minimal rdf with weights $w_{f_1}=2$, and $w_{f_2}=w_{f_3}=n+1$. 

In~\cite{AbuFerMan2022},  several combinatorial properties of minimal Roman dominating functions
were derived that were central for obtaining a general algorithmic enumeration result and that are also important when studying special graph classes.
This is summarized as follows.

\begin{theorem}\label{thm:property_min_rdf}
Let $G=\left(V,E\right)$ be a graph, $f: \: V \rightarrow \lbrace 0,1,2\rbrace$ and abbreviate
$G'\coloneqq G\left[ V_0\left(f\right)\cup V_2\left(f\right)\right]$. Then, $f$ is a minimal rdf if and only if the following conditions hold:
\begin{enumerate}
\item$N_G\left[V_2\left(f\right)\right]\cap V_1\left(f\right)=\emptyset$,\label{con_1_2}
\item $\forall v\in V_2\left(f\right) :\: P_{G',V_2\left(f\right)}\left( v \right) \nsubseteq \lbrace v\rbrace$, also called \emph{privacy condition}, and \label{con_private}
\item $V_2\left(f\right)$ is a minimal dominating set of $G'$.\label{con_min_dom}
\end{enumerate}
\end{theorem}

This combinatorial result has been the key to show a polynomial-time decision procedure for the extension problem (concerning a given function $f:V\to\{0,1,2\}$). But it can also be used to design  enumeration algorithms that are also input-sensitive.
The simplest exploit is to branch on all vertices whether or not a vertex should belong to $V_2(f)$.
Once $V_2(f)$ is fixed, its neighborhood will form $V_0(f)$ and the remaining vertices will be $V_1(f)$.
To achieve better running times, this approach has to be clearly refined. For more details, see \autoref{sec:general}.

\paragraph*{Outline of the presentation}
As a warm-up, we consider enumerating minimal rdf in 
split and cobipartite graphs, for which we present a very simple algorithm that does not make use of any more advanced techniques like Measure-and-Conquer.
Then, we turn our attention to (exact) counting of minimal rdf on paths, an exercise that will turn out to be an important step for branching algorithms for forests and for interval graphs. In all these cases, we obtain enumeration algorithms that are provable optimal in the sense that there are possible input graphs that require the number of minimal rdf to be output as proved as an upper bound for the respective families of graphs.
We conclude the paper with discussing the family of chordal graphs, where we present an enumeration algorithm with a running time substantially better than in the general case, although we do not know if the upper bound is tight.

\section{Enumerating Minimal Roman Dominating Functions in Split and in Cobipartite Graphs}

A split graph $G=(V,E)$ consists of a bipartition of $V$ as $C$ and $I$, such that~$C$ forms a clique and~$I$ is an independent set. Let $f:V\to\{0,1,2\}$ be a minimal rdf of $G$. Then, if $V_2=f^{-1}(2)$ contains both a vertex~$v_c$ from $C$ and a vertex~$v_i$ from~$I$, then $v_i$ cannot find a private neighbor  in~$G$, contradicting minimality of~$f$. We can hence first branch to decide if  $V_2\subseteq C$ or if $V_2\subseteq I$.
After dealing with the simple case that $|V_2\cap C|=1$ separately, we can assume that all private neighbors of $V_2\subseteq C$ are in~$I$
 and that all private neighbors of $V_2\subseteq I$ are in~$C$. We will describe a simple branching algorithm in which we can assume to immediately delete vertices that are assigned the value~$0$, as they will be always dominated.
 
\smallskip\noindent
\underline{Case 1.} One element of $C$ is assigned a value of~$2$. We can guess this element in $\mathcal{O}(n)$ and proceed as follows.
\\
1. Elements of $C$ with no neighbors in $I$ are assigned a value of zero.
\\
2. Pick $v \in C$ with at least two neighbors in $I$ and branch by either setting $f(v)=2$ and assign~$0$ to vertices in $N(v)\cap I$ or $f(v)=0$ (this leads to the branching vector $(3,1)$).
\\
3. When all elements of $C$ have exactly one neighbor in $I$, 
pick some $v\in C$ with $N(v)\cap I=\{w\}$. Distinguish two cases.\\
3.1 $w$ has at least one other neighbor $x\in C$. Then either $f(v)=2$, $f(w)=f(x)=0$ (in fact, all neighbors of $w$ are assigned~0), or $f(v)=0$ (this leads to a $(3,1)$ branch).\\
3.2 $N(w) = \{v\}$: either $f(v)=2$, $f(w)=0$ or $f(v)=0$, $f(w)=1$ (this leads to the branching vector $(2,2)$).

\smallskip
\noindent
\underline{Case 2.} No element of $C$ is assigned a value of~$2$.\\
1. 
Then any isolated element of $I$ is automatically assigned a value of 1 and can be deleted. Moreover, any element of $C$ with no neighbors in $I$ is assigned a value of 1 and deleted.
\\
2. Pick a vertex $v$ of degree at least two in $I$ and branch by either setting $f(v)=2$ and assigning zero to all its neighbors or set $f(v)=1$ (this leads to the branching vector $(3,1)$).
\\
3. When all elements of $I$ are pendant, pick $v\in I$ with $N(v)\cap C=\{w\}$. Distinguish 2 cases.\\
3.1 $w$ has at least one more neighbor $x\in I$: either $f(v)=2,f(w)=0,f(x)=1$ or $f(v)=1$ (delete $v$) (this leads to the branching vector $(3,1)$).\\
3.2 $N(w)\cap I = \{v\}$: either $f(v)=2,f(w)=0$ or $f(v)=f(w)=1$ (this leads to the branching vector $(2,2)$).

The worst-case branch vector is $(1,3)$, which leads to the following claim.
\begin{theorem}
All minimal rdf in a split graph of order~$n$ can be enumerated in time $\Oh^*(\RomanSplitUpperbound^n)$.
\end{theorem}

For cobipartite graphs, a similar reasoning applies.
Now, it could be possible that one vertex~$x$ of the bipartition side $X$ finds its private neighbor~$p_x$ in~$X$ itself and that one vertex~$y$ of the other bipartition side $Y$ finds its private neighbor~$p_y$ in~$Y$, such that the edges $xp_y$ and $yp_x$ do not exist. If $G$ contains no universal vertices, then irrespectively whether the $V_2$-vertices lie in $X$ or in $Y$, there must be at least one other vertex in $V_2$ on the same side.
But this means that they must find their private neighbors on the other side. The branching is hence analogous to the split graph case.
 \begin{theorem}
All minimal rdf in a cobipartite graph of order~$n$ can be enumerated in time $\Oh^*(\RomanSplitUpperbound^n)$.
\end{theorem}

The previous arguments are not valid in the case of bipartite graphs.
Here, we rather suspect that the general case is not really easier than the bipartite case.

For the lower bound for split graphs (cobipartite graphs have a have similar example) we look at the graph $G=(C \cup I, E)$ with $C=\{c_1,\ldots,c_n\}$, $I=\{v_1,\ldots,v_n\}$ and $E=\binom{C}{2}\cup\{\{c_1,v_1\},\ldots,\{c_n,v_n\}\}$ (for the cobipartite case $I$ is also a clique). For $c_i\in V_2\subseteq C$, $v_i$ would be a private neighbor and $c_i$ would be a private neighbor for $v_i\in V_2\subseteq I$. Thus, each subset of $C$ and each subset of $I$ is a possible choice for $V_2$. Therefore, $G$ has $2\cdot\sqrt{2}^{n}-1$ minimal rdf (for the cobipartite case there are $2\cdot\sqrt{2}^{n}+\frac{n^2}{4}-1$ possibilities, since we could also choose any vertex of $C$ together with any vertex of $I$).

\begin{corollary}
There exist split and cobipartite graphs of order~$n$ with $\Omega(\sqrt{2}^n)$ many minimal rdf.
\end{corollary}

\section{Counting Minimal Roman Dominating Functions on 
Paths} 
\label{sec:RomanPaths}
In this section, we will develop formulas for the number of minimal Roman dominating functions on paths.

Let $C_{P,n}$ count the number of minimal rdf of a $P_n$. Furthermore, let
$C_{P,{2},n}$ and $C_{P,\overline{2},n}$ denote the number of minimal rdf  of a $P_n$ where the first vertex is assigned 2, or where it is decided that the first vertex is not assigned 2, respectively.
Clearly, 
$$C_{P,n}=C_{P,{2},n}+C_{P,\overline{2},n}\,.$$
Consider  $P_n=\left(V_n,E_n\right)$ with $V_n=\{v_i\mid i\in [n]\}$ and $E_n=\{v_iv_{i+1}\mid i\in [n-1]\}$.
Let $n\geq 3$ and $f:V_n\to\{0,1,2\}$ be a minimal rdf. If $f(v_1)=2$, then $f(v_2)=0$ and it is clear that $f(v_3)\neq 2$. This shows (including trivial initial cases):
$$C_{P,{2},n}=\begin{cases}
0,&\text{if }n=1\\
1,&\text{if }n=2\\
C_{P,\overline{2},n-2},&\text{if }n>2
\end{cases}$$
If $f(v_1)\neq 2$, then we have two subcases: (a) if $f(v_1)=1$, then we know that $f(v_2)\neq 2$;
(b) if $f(v_1)=0$, then $f(v_2)=2$ is enforced. But we know a bit more compared to the initial situation: This 2 on $v_2$ has already a private neighbor, namely $v_1$. Therefore, we have more possibilities for $v_3$: either $f(v_3)=0$ or $f(v_3)=2$. The second subcase is as before, because this 2 has no private neighbor.
If  $f(v_3)=0$, then either $f(v_4)=2$, and this 2 has no private neighbor, or $f(v_4)\neq 2$.
Hence, we find the recursion established in \autoref{fig:CPnot2n-formula}.

\begin{figure}
   $$C_{P,\overline{2},n}=\begin{cases}
1,&\text{if }n=1\\ 
2,&\text{if }n=2\\
3,&\text{if }n=3\\
C_{P,\overline{2},n-1}+C_{P,{2},n-2}+\overbrace{C_{P,{2},n-3}+C_{P,\overline{2},n-3}}^{=C_{P,n-3}},&\text{if }n>3
\end{cases}$$
    \caption{Recursion formula for minimal rdf of $P_n$ not starting with the label~2.}
    \label{fig:CPnot2n-formula}
\end{figure}

Now, $C_{P,n}=C_{P,{2},n}+C_{P,\overline{2},n}=C_{P,\overline{2},n-2}+C_{P,\overline{2},n-1}+C_{P,{2},n-2}+C_{P,n-3}=C_{P,\overline{2},n-1}+C_{P,n-2}+C_{P,n-3}$. Conversely,  $C_{P,n}=C_{P,{2},n}+C_{P,\overline{2},n}=C_{P,\overline{2},n-2}+C_{P,\overline{2},n}.$
Hence, $C_{P,n}=C_{P,\overline{2},n}+C_{P,\overline{2},n-2}=C_{P,\overline{2},n-1}+(C_{P,\overline{2},n-2}+C_{P,\overline{2},n-4})+(C_{P,\overline{2},n-3}+C_{P,\overline{2},n-5})$, which gives, ignoring the cases for small values of $n$:
$$C_{P,\overline{2},n}=C_{P,\overline{2},n-1}+C_{P,\overline{2},n-3}+C_{P,\overline{2},n-4}+C_{P,\overline{2},n-5}\approx 1.6852^n$$
As $C_{P,n}=C_{P,\overline{2},n-2}+C_{P,\overline{2},n}$, the same asymptotic behavior holds for $C_{P,n}$, i.e., $C_{P,n}=\mathcal{O}^*(1.6852^n)$.

\begin{proposition}
The number of minimal Roman dominating functions of a path $P_n$ grows as $\mathcal{O}^*(c_{\text{RD,P}}^n)$, with $c_{\text{RD,P}}\leq 1.6852$.
\end{proposition}

This should be compared with the recursion of Br\'od~\cite{Bro2011} that yields the following asymptotic behavior for the number of minimal dominating sets of a path with $n$ vertices:
\begin{corollary} \cite{Bro2011}
The number of minimal dominating sets of a path $P_n$ grows as $\mathcal{O}^*(c_{\text{D,P}}^n)$, with $c_{\text{D,P}}\leq 1.4013$.
\end{corollary}
As every minimal dominating set $D\subseteq V$ of a graph $G=(V,E)$ corresponds to the minimal rdf $f:V\to\{0,1,2\}$ with $f^{-1}(2)=D$ and $f^{-1}(0)=V\setminus D$, it is clear that $c_{\text{D,P}}\leq c_{\text{RD,P}}$.\footnote{While the sequence of the numbers of minimal dominating sets of a $P_n$ can be found in
the Encyclopedia of Integer Sequences, this is not the case for  the sequence of the numbers of minimal rdf of a $P_n$.}

Apart from bringing insights into the number of minimal rdf of single paths of a certain length, our recursions are also helpful to determine 
how many minimal rdf can be in forests of paths. The interesting corner cases are here determined by the graph families $\mathcal{P}_n$ that consists of the graph union of arbitrary many paths on~$n$ vertices, such that the order of each graph $G\in \mathcal{P}_n$ is divisible by~$n$.
\autoref{tab:forests-of-paths} shows the number of minimal rdf per connected component (i.e., it spells out $C_{P,n}$) and the and branching numbers.

\begin{table}[tbh]
    \centering
    \begin{tabular}{c||c|c|c|c|c|c|c}
    $n$ & 1 & 2 & 3 & 4 & 5 & 6 & 7\\\hline 
    $C_{P,n}$ & 1 & 3 & 4 & 7 & 12 & 20 & 34\\
    branching number & 
    1 & $\sqrt{3}\leq 1.7321$ & 1.5875 & 1.6266 & 1.6438 & 1.6476 & 1.6550\\
\end{tabular}
    \caption{Branching numbers for collections of paths}
    \label{tab:forests-of-paths}
\end{table}

Hence, we can conclude:

\begin{corollary}
Within the family of forests of paths, the number of minimal rdf for graphs of order~$n$ is $\Oh(\RomanIntervalExact^n)$, a value actually approached by $\mathcal{P}_2$.\footnote{We have a cumbersome argument that the intuition that the exponential growth will `win' against the polynomial factors from~3 onwards.}
\end{corollary}

We will further extend this result towards forests and towards interval graphs in the next sections, starting with a more general description of such  branching algorithms for enumerating minimal rdf. 

\section{A General Approach to Branching for Minimal RDF} 
\label{sec:general}

In this section, we sketch the general strategy that we apply for enumerating minimal rdf.
In most cases, the branching will look for a yet undecided vertex~$v$ (that we will call \emph{active} henceforth) and will decide to label it with~$2$ in one branch and not to label it with~$2$ in the other branch. Now, in the first branch, we can say something about the neighbors of~$v$ as well: according to \autoref{thm:property_min_rdf}, they cannot be finally labelled with~$1$. We express this and similar properties by (always) splitting the vertex set $V$ of the current graph $G=(V,E)$ into:

\begin{itemize}

\item[-] $\Active$: notice that in the very beginning of the branching, all vertices are active.
%set of active vertices, i.e., no assigned value/restriction.

\item[-] $\overline{V_i}$: vertices that cannot be assigned a value of $i$, $i\in\{1,2\}$.

\item[-] $V_0$: set of vertices assigned a value of zero that are not yet dominated.

\end{itemize}

Sometimes, the branching also considers a vertex from $\overline{V_1}$, which will be assigned~$0$ (and hence is deleted) in the branch when it is not assigned~$2$. We can also call extendibility tests before doing the branching in order to achieve polynomial delay; see \cite{AbuFerMan2022}.

Possibly, we can also (temporarily) have (and speak of) vertex sets $V_i$ (with $i\in\{1,2\}$) with the meaning that each vertex in $V_i$ is assigned the value~$i$. Our algorithms will preserve the invariant that a vertex $v\in\overline{V_1}$ must have a  neighbor put into $V_2$ (in the original graph). However, notice that once the effect (mostly implied by \autoref{thm:property_min_rdf}) of putting a vertex~$v$ into $V_i$ on its neighborhood $N(v)$ has been taken care of, these vertices can be deleted from the `current graph' to simplify the considerations. More precisely, for $i\in \{1,2\}$, our algorithms automatically delete vertices assigned a value of $i$ after making sure the neighbors are placed in $\overline{V_{3-i}}$.
It could happen that the neighbor of a vertex~$w\in\overline{V_2}$ is assigned the value~$2$. Then, $w$ must be assigned~$0$; as it is dominated, it can and will be deleted.
Similarly, if  the neighbor of a vertex~$w\in\overline{V_1}$ is assigned the value~$1$, $w$ must be assigned~$0$ and is hence deleted. 
Only finally, it should be checked if a function $f:V\to \{0,1,2\}$ that is constructed during branching is indeed a minimal rdf, because possibly some vertices assigned~$2$ do not have a private neighbor.
During the course of our algorithm, whenever we speak of the degree of a vertex (in the current graph) in the following, we only count in neighbors in $\overline{V_1}\cup\overline{V_2}\cup \Active$.
We sometimes abbreviate $V_0 \cup \overline{V_2}( \cup V_1)$ as  $\widehat{V_2}$. 

Reduction rules are an important ingredient of any branching algorithm. 
We will make use of the following reduction rules.
Rules in similar form appeared in~\cite{AbuFerMan2022}. 

\begin{redrule}\label{not-V2-with not-V2-neighbors}
If $v \in \overline{V_2}$ with $N(v) \subseteq \overline{V_2}%\cup V_1\cup V_0
$, then set $f(v)=1$ and delete $v$.
\end{redrule}

\begin{redrule}\label{isolated-not-in-V1}
If $v \in \overline{V_1}$ with $N(v) \subseteq \overline{V_1}%\cup V_2\cup V_0
$, then set $f(v)=0$ and delete $v$.
\end{redrule}

\begin{redrule}\label{all-neighbors-not-in-V1}
If $v\in \Active$ obeys $N(v)\subseteq \overline{V_1}$, then put $v$ into $\overline{V_2}$.
\end{redrule}

\begin{lemma}
The three reduction rules are sound.
\end{lemma}

\begin{pf} First consider $v \in \overline{V_2}$ with $N(v) \subseteq \overline{V_2}$.
If $v$ would be assigned~$0$, then it would need a neighbor assigned~$2$ to dominate it. Therefore, only the possibility to assign~$1$ to~$v$ remains. Secondly, consider $v \in\Active\cup  \overline{V_1}$ with $N(v) \subseteq \overline{V_1}$. If $v$ would be assigned~$2$, then $v$ must need a private neighbor in $N(v)$. As all of $N(v)$ is part of $\overline{V_1}$, by our invariant this means that all these vertices are already dominated. Therefore, $v$ cannot be assigned~$2$. 
\end{pf}

Similarly as in~\cite{AbuFerMan2022},
we will perform a Measure-and-Conquer analysis of the branching algorithms that we will describe.
As a measure, we take $$\mu(\Active,\overline{V_1},\overline{V_2},V_0) = |\Active| + \omega_1\, |\overline{V_1}| + \omega_2\, | \overline{V_2}|$$
for the `current graph' with vertex set $\Active\cup \overline{V_1}\cup \overline{V_2}\cup V_0$.
In the beginning of the algorithm, $\Active=V$ and $\overline{V_1}= \overline{V_2}= V_0=\emptyset$. 
To explain the work of the reduction rules, consider an isolated vertex (in the very beginning). The reduction rules will first move it into $\overline{V_2}$ and then into $V_1$ to finally delete it.

We will choose the constants $\omega_1,\omega_2\in [0,1]$ to assess the running times of our algorithms best possible, hence also delivering upper bounds on the number of minimal rdf of graphs of order~$n$ belonging to a specific graph class.

Concerning the reduction rules, we can easily observe that their application will never increase the measure.
We will list in the following several branching rules (for the different graph classes) and we always assume that the rules are carried out in the given order.

\section{Enumerating  Minimal RDF %Roman Dominating Functions 
on Interval Graphs}\label{sec:min_rdf_interval}

Recall that an \emph{interval graph} can be described as the intersection graph of a collection of intervals on the real line. This means that the vertices correspond to intervals and that there is an edge between two such vertices if the intervals have a non-empty intersection.
We assume in the following that $G= (V,E)$ is a interval graph with  the interval representation $\mathcal{I}=\{I_v:= [l_v, r_v]\}_{v\in V}$, i.e., $l_v$ is the left border and $r_v$ is the right border of the interval representing the vertex~$v$. We call $v\in U$ \emph{leftmost in $U$} if it is a vertex from $U$ that has the smallest value of $r_u$ among all vertices in~$U$. A vertex leftmost in~$V$ is simply called leftmost.

As we are rather dealing with the partition of the vertex set of the current graph into $\Active,\overline{V_1},\widehat{V_2}$ in the following branching algorithm for interval graphs, we re-formulate Reduction Rules~\ref{isolated-not-in-V1} and~\ref{all-neighbors-not-in-V1} as one rule:

\begin{redrule}\label{RRNoNeighbor}
Let $v \in \Active \cup \overline{V_1}$ with $N(v) \subseteq \overline{V_1} \cup (\widehat{V_2}\setminus \overline{V_2})$. Then put $v$ into~$\widehat{V_2}$. 
\end{redrule}

\noindent
\autoref{RRNoNeighbor} implies that each vertex in $v\in A$ has at least one neighbor in $A\cup V_2$. Concerning the measure, we will have $\omega_1=1$ and set $\omega_2=\omega=0.57$.
We are now going to present the branching rules that constitute the backbone of our algorithm for enumerating minimal rdf on interval graphs. For the convenience of the reader, we also provide illustrations of the different branching scenarios. In these figures, we adhere to the following drawing conventions:
\begin{itemize}
    \item \tikz{\draw[circle,fill=white, scale=0.8]  circle(1ex);} are vertices in $\Active$.
    \item \tikz{\draw[circle,fill=black!40, scale=0.8]  circle(1ex);} are vertices in $\overline{V_1}$
    \item \textbf{\tikz{\draw[rectangle,fill=white,scale=0.8] (0,0) rectangle (2ex,2ex);}} are vertices in $\overline{V_2}$
    \item \tikz{\draw (0,0) node[diamond,scale=0.7, draw] {};} are vertices in $A\cup \overline{V_2}$, for which the exact set is not further defined.
    \item \tikz[fill lower half/.style={path picture={\fill[#1] (path picture bounding box.south west) rectangle (path picture bounding box.east);}}]{\draw (0,0) node[diamond,fill lower half=black!40,scale=0.7, draw] {};} are vertices in $A\cup \overline{V_1} \cup \overline{V_2}$, for which the exact set is not further defined. 
\end{itemize}

\begin{brarule}\label{BRDom}
Let $v$ be the leftmost vertex in $\overline{V_1}$ and let $u$ be the leftmost vertex in $N(v) \cap (\Active \cup \overline{V_2})$ and branch as follows:

\begin{enumerate}
    \item Put $v$ in $\widehat{V_2}$.
    \item Put $v$ in $V_2$ and $u \in \widehat{V_2}$ (if it is not there yet).
\end{enumerate}
\end{brarule}

\begin{lemma}\label{lem:BRDom}
The branching is a complete case distinction. Moreover, it leads at least to the following branching vector:
$(1, 1 + \omega)\, .$
\end{lemma}

\begin{pf}
Assume there exists a minimal rdf $f$ with $V_2(f) \cap (V_2 \cup \widehat{V_2})= V_2$.
To prove that this is a complete case distinction, we show: if $v\in V_2(f)$ holds, $u \in V_0(f)$ is the private neighbor of~$v$. Let $w \in V_2 \cap N[v]$. By \autoref{RRNoNeighbor}, $N[v]\setminus N[w]$ is not empty. By the construction of the algorithm (look at the proof of \autoref{mainInterval}, marked $(*)$) and the minimality of~$f$ ($I_v\nsubseteq I_w$), $l_w< l_v < r_w < r_v$ holds.
\\
\underline{Case 1:} $I_u \subseteq I_v$.
Thus $f(u)\neq 2$. If $u$ is not a private neighbor of~$v$, then there exists a $t\in V_2(f)\cap N(v)$. Because of the minimality of $f$, $l_v < l_t < r_u < r_v < r_t$ holds. Since $v$ needs a private neighbor, there exists $p\in N(v)\setminus N[\{w,t\}]$. This implies $I_p\cap I_v \subseteq I_v\setminus \{ I_w \cup I_t\}=(r_w, l_t)$. This is a contradiction to the minimality of~$u$.
\\
\underline{Case 2:} $I_u \nsubseteq I_v$.
Since $u$ is not dominated by $V_2$, we have $l_w < l_v < r_w < l_u < r_v < r_u$. Assume there exists a $t\in N[u] \cap V_2(f)$. By minimality of $f$, $I_t\nsubseteq I_v$, $I_v\nsubseteq I_w \cup I_t$ and there exists a $p \in N(v)\setminus N[\{ w, t\}]$. This implies $I_p\cap I_v \subseteq I_v\setminus ( I_w \cup I_t)=(r_w, l_t)\subseteq (r_w, r_u)$. This contradicts the minimality of~$r_u$.

Thus, $u$ has to be the private neighbor of $v$. Therefore, the measure is decreased by $1+\omega$, if $v\in V_2$. Since $v$ is already dominated, it would move into $\widehat{V_2}\setminus \overline{V_2}$ if it is put into $\widehat{V_2}$.  This decreases the measure by one. 
\end{pf}

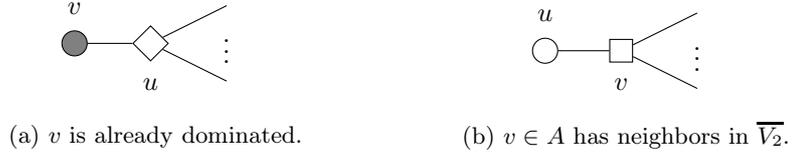
\begin{figure}[bt]
    \centering
    	
\begin{subfigure}[t]{.4\textwidth}
    \centering
    	
	\begin{tikzpicture}[transform shape]
			\tikzset{every node/.style={ fill = black,circle,minimum size=0.3cm}}
			\node[draw,fill=gray,label={above:$v$}] (u) at (0,0) {};
			\node[draw,diamond,fill=white,label={below:$u$}] (v1) at (1,0) {};
			\node[fill=none] at (2,0) {$\vdots$};
			\path (u) edge[-] (v1);
			\path (v1) edge (2,-0.5);
			\path (v1) edge (2,0.5);

        \end{tikzpicture}

    \subcaption{%Branching Rule~\ref{branching:dom-nbs-On}: 
    $v$ is already dominated.}
    \label{fig:BRDom}
\end{subfigure}
\qquad
\begin{subfigure}[t]{.48\textwidth}
    \centering
    	
	\begin{tikzpicture}[transform shape]
			\tikzset{every node/.style={ fill = black,circle,minimum size=0.3cm}}
			\node[draw,fill=white,label={above:$u$}] (u) at (0,0) {};
			\node[draw,rectangle,fill=white,label={below:$v$}] (v1) at (1,0) {};
			\node[fill=none] at (2,0) {$\vdots$};
			\path (u) edge[-] (v1);
			\path (v1) edge (2,-0.5);
			\path (v1) edge (2,0.5);

        \end{tikzpicture}

    \subcaption{%Branching Rule~\ref{branching:dom-nbs-On}: 
    $v\in \Active$ has neighbors in $\overline{V_2}$.}
    \label{fig:BRNotDominatable}

    \end{subfigure}
    \caption{Branching Rules~\ref{BRDom}  and~\ref{BRNotDominatable}}
\end{figure}
\begin{brarule}\label{BRNotDominatable}
Let $v$ be the leftmost vertex in $(\Active \cup \overline{V_2})$. If $v\in \Active$ and $N(v) \cap \Active = \emptyset$ hold, branch as follows:

\begin{enumerate}
    \item Put $v$ in $V_2$.
    \item Put $v$ in $\widehat{V_2}$.
\end{enumerate}
\end{brarule}

%\begin{toappendix}
\begin{lemma}
The branching is a complete case distinction. Moreover, it leads at least to the following branching vector:
$( 1 + \omega, 1)\, .$
\end{lemma}

\begin{pf}
It is clear that this is a complete case distinction. Because of  \autoref{RRNoNeighbor}, we get $ N(v) \cap \overline{V_2} \neq \emptyset$. Assume there exists a minimal rdf $f$ with $V_2(f) \cap (V_2 \cup \widehat{V_2})= V_2$. If $f(v)=2$, the vertices in $N(v) \cap \overline{V_2}$ would be 
dominated now. This decreases the measure by at least $1 + \omega$.
If $f(v) \neq 2$, it would be in $\widehat{V_2} \setminus \overline{V_2}$, as it has no neighbor in $\Active$. Thus, the measure is decreased by at least~$1$.
\end{pf}
%\end{toappendix}

\begin{brarule}\label{BRP0}
Let $v$ be leftmost in $(\Active \cup \overline{V_2})$. If $v\in \Active$ and $\vert N(v) \cap \Active \vert \geq 2$ hold, branch as follows:

\begin{enumerate}
    \item Put $v$ in $V_2$ and all vertices in  $N(v) \cap \Active$ into $\widehat{V_2}$.
    \item Put $v$ in $\widehat{V_2}$.
\end{enumerate}
\end{brarule}

%\begin{toappendix}
\begin{lemma}\label{LemBRP0}
The branching is a complete case distinction. Moreover, it leads  to a branching vector not worse than
$(3, 1 - \omega)\, .$
\end{lemma}

\begin{pf}
Assume there exists a minimal rdf $f$ with $V_2(f) \cap (V_2 \cup \widehat{V_2}) = V_2$. The second branch is clear (also concerning the measure). For the first branch, we assume there exists a $u\in N(v)\cup \Active$ with $f(u) = f(v) = 2$. Since both need a private neighbor, $I_v \nsubseteq I_u$ holds. This implies $l_v < l_u < r_v < r_u$. Let $y\in N(v) \cap (\Active \cup V_2)$ be the private neighbor of $v$. Therefore, $I_v\cap I_y \subseteq I_v \setminus I_u = [l_v,l_u)$ holds. Thus $r_y < r_v$. %\todo{changed to from $r_u$ to $r_y$} 
This would contradict the minimality of $v$. This implies that this branching is a complete case distinction. Since we lose the weight of three vertices in $\Active$, all from  $N[v]\cap \Active$, the measure decreases by at least~$3$.
\end{pf}

%\end{toappendix}

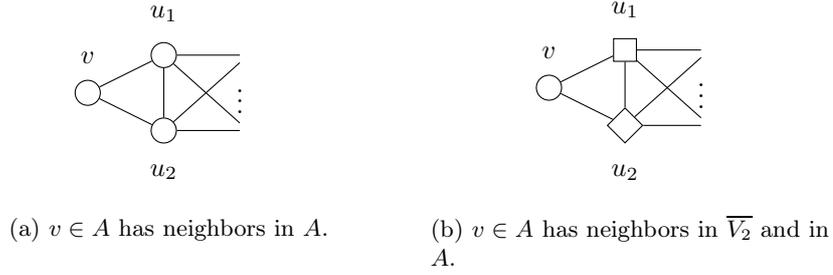
\begin{figure}[bt]
    \centering
    	
\begin{subfigure}[t]{.43\textwidth}
    \centering
	\begin{tikzpicture}[transform shape]
			\tikzset{every node/.style={ fill = white,circle,minimum size=0.3cm}}
			\node[draw,label={above:$v$}] (v) at (0,0) {};
			\node[draw,label={above:$u_1$}] (u1) at (1,0.5) {};
			\node[draw,label={below:$u_2$}] (u2) at (1,-0.5) {};
			\node[fill=none] at (2,0) {$\vdots$};
			\path (u1) edge[-] (v);
			\path (u2) edge[-] (v);
		    \path (u1) edge[-] (u2);
			\path (u2) edge (2,0.4);
			\path (u1) edge (2,-0.4);
			\path (u2) edge (2,-0.5);
			\path (u1) edge (2,0.5);
        \end{tikzpicture}

    \subcaption{%Branching Rule~\ref{branching:dom-nbs-On}: 
    $v\in \Active$ has neighbors in $\Active$.}
    \label{fig:BRP0}
    \end{subfigure}
\qquad
\begin{subfigure}[t]{.43\textwidth}
    \centering
    	
	\begin{tikzpicture}[transform shape]
		\tikzset{every node/.style={ fill = white,circle,minimum size=0.3cm}}
			\node[draw,label={above:$v$}] (v) at (0,0) {};
			\node[draw,rectangle,label={above:$u_1$}] (u1) at (1,0.5) {};
			\node[draw,diamond,label={below:$u_2$}] (u2) at (1,-0.5) {};
			\node[fill=none] at (2,0) {$\vdots$};
			\path (u1) edge[-] (v);
			\path (u2) edge[-] (v);
		    \path (u1) edge[-] (u2);
			\path (u2) edge (2,0.4);
			\path (u1) edge (2,-0.4);
			\path (u2) edge (2,-0.5);
			\path (u1) edge (2,0.5);

        \end{tikzpicture}

    \subcaption{%Branching Rule~\ref{branching:dom-nbs-On}: 
    $v\in \Active$ has neighbors in $\overline{V_2}$ and in $\Active$.}
    \label{fig:BR_P2TildeV2}
\end{subfigure}

    \caption{Branching Rules~\ref{BRP0}  and~\ref{BR_P2TildeV2}}
\end{figure}

\begin{brarule}\label{BR_P2TildeV2}
Let $v$ be the leftmost vertex in $(\Active \cup \overline{V_2})$. If $v\in \Active$, $\vert N(v) \cap \overline{V_2} \vert \geq 1$ and $\vert N(v) \cap \Active \vert =1$ with $u\in N(v) \cap \Active$ hold, then branch as follows:

\begin{enumerate}
    \item Put $v$ in $V_2$ and $u$ into $\widehat{V_2}$.
    \item Put $u$ in $V_2$ and $v$ in $\widehat{V_2}$.
    \item Put $u,v$ in $\widehat{V_2}$.
\end{enumerate}
\end{brarule}

%\begin{toappendix}

\begin{lemma}
The branching is a complete case distinction. Moreover, it leads to a branching vector not worse than
$(2 + \omega, 2 + \omega, 2 - \omega)\,. $
\end{lemma}

\begin{pf} First we show that $N(v) \cap (\Active \cup \overline{V_2})$ is a clique. Let $x\in N(v) \cap (\Active \cup \overline{V_2})$. Thus, we  find $r_v < r_x$ and $I_v \cap I_x \neq \emptyset$. Hence, there exists a $t\in I_v \cap I_x$ with $ t < r_v$, so that $[t,r_v]\subseteq I_x$  holds. This implies $r_v\in I_x$ for each $x \in N(v) \cap (\Active \cup \overline{V_2})$ and $N(v) \cap (\Active \cup \overline{V_2})$ is a clique. 

Since $v$ is simplical on $G\left[\Active \cup \overline{V_2}\right]$, $N_{G\left[\Active \cup \overline{V_2}\right]}(v)\subseteq N_{G\left[\Active \cup \overline{V_2}\right]}(u)$ holds. Thus, for $v,u\in V_2$, $v$ has no private neighbor. This contradict the minimality of a minimal rdf.  

In the first branch, the measure decreases by at least $2 + \omega$, since $v$ gets into~$V_2$, $u$ gets into $\widehat{V_2}$ and all vertices in $N(v)$ are dominated. 

As $N[v]\cap (\Active \cup \overline{V_2})$ is a clique, it is dominated by $u$ in the second branch and the measure is decreasing by $2 + \omega$.

In the third branch, we put $v$ into $\widehat{V_2}\setminus \overline{V_2}$, since there is no vertex in $N[v]\cap \Active$ anymore. Furthermore, $u$ moves into $\widehat{V_2}$. Thus, this case decreases the measure by $2-\omega$.  
\end{pf}

\begin{brarule}\label{BR_P1}
Let $v$ be the leftmost vertex in $(\Active \cup \overline{V_2})$. If $ N[v]\cap \Active  = \{u\}$ with $N[v] \cap \overline{V_2} = \emptyset$ and $\vert N(u) \cap \Active\vert \geq 3$, branch as follows:

\begin{enumerate}
    \item Put $v$ in $V_2$ and $N[u]\setminus \{ v \}$ in $\widehat{V_2}$.
    \item Put $v$ in $\widehat{V_2}$.
\end{enumerate}
\end{brarule}

%\begin{toappendix}
\begin{lemma}
The branching is a complete case distinction. Moreover, it leads to a branching vector not worse than $( 2 + 2\cdot (1-\omega), 1 - \omega)\, .$
\end{lemma}

\begin{pf}Consider vertices $u,v$ as described in the branching rule.
Let $f: V \to \{0,1,2\}$ be a minimal rdf with $V_2 \subseteq V_2(f)$ and $\widehat{V_2}\cap V_2(f) = \emptyset$. If $v\in V_2(f)$, then $u$ has to be the private neighbor of $v$. This implies $\left(N[u]\setminus \{v\}\right)\cap V_2(f) \neq \emptyset$. Therefore, the the measure decreases by $2 + (\vert N(u) \cap \Active\vert - 1)\cdot (1-\omega)\geq 2+2\cdot (1-\omega)$. The other case is trivial.
\end{pf}
%\end{toappendix}

\begin{figure}[bt]
    \centering
\begin{subfigure}[t]{.42\textwidth}
    \centering
	\begin{tikzpicture}[transform shape]
			\tikzset{every node/.style={ fill = white,circle,minimum size=0.3cm}}
			\node[draw,label={above:$v_1$}] (v1) at (0,0) {};
			\node[draw,label={above:$v_2$}] (v2) at (1,0) {};
			\node[draw,label={below:$u_2$}] (u2) at (2,-0.5) {};
			\node[draw,label={above:$u_1$}] (u1) at (2,0.5) {};
			\node[fill=none] at (3,0) {$\vdots$};
			\path (v1) edge[-] (v2);
			\path (u1) edge[-] (v2);
			\path (u2) edge[-] (v2);
		    \path (u1) edge[-] (u2);
			\path (u2) edge (3,0.4);
			\path (u1) edge (3,-0.4);
			\path (u2) edge (3,-0.5);
			\path (u1) edge (3,0.5);
        \end{tikzpicture}
    \subcaption{%Branching Rule~\ref{branching:dom-nbs-On}: 
    $v_1\in \Active$ has only on neighbor in~$\Active$ which has degree bigger than~2.}
    \label{fig:BRP1}
    \end{subfigure}
    \qquad
\begin{subfigure}[t]{.48\textwidth}
    \centering
    	
	\begin{tikzpicture}[transform shape]
		\tikzset{every node/.style={ fill = white,circle,minimum size=0.3cm}}
			\node[draw,label={above:$v_1$}] (v1) at (0,0) {};
			\node[draw,label={above:$v_2$}] (v2) at (1,0) {};
			\node[draw,label={below:$v_3$}] (v3) at (2,0) {};
			\node[draw,label={above:$u$}] (u) at (2,1) {};
			\node[fill=none] at (3,0) {$\vdots$};
			\path (v1) edge[-] (v2);
			\path (v2) edge[-] (v3);
		    \path (v3) edge[-] (u);
			\path (v3) edge (3,0.5);
			\path (v3) edge (3,-0.5);

        \end{tikzpicture}

    \subcaption{%Branching Rule~\ref{branching:dom-nbs-On}: 
    $v_1,v_2,v_3\in \Active$ is a path and $v_3$ has a pendant neighbor.}
    \label{fig:BR_P2single}
\end{subfigure}

    \caption{Branching Rules~\ref{BR_P1}  and~\ref{BR_P2single}}
\end{figure}
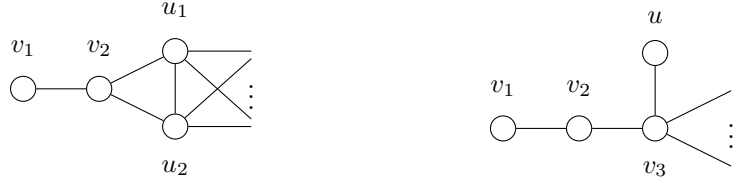
%\todo{Perhaps we have to show, that there is no possibility $u$ has a neighbor in $\overline{V_2}$}

\begin{brarule}\label{BR_P2single}
Let $v_1$  be the leftmost vertex in $(\Active \cup \overline{V_2})$. If $ N[v_1]\cap \Active  = \{v_2\}$ with $N[v_1] \cap \overline{V_2} = \emptyset$, $ N(v_2) \cap \Active =\{v_1,v_3\}$ and if there exists a  $u \in N(v_3)$ such that $N(u)= \{ v_3 \}$, then branch as follows:

\begin{enumerate}
    \item Put $v_1$ into $V_2$ and $v_2,v_3$ into~$\widehat{V_2}$.
    \item Put $v_1,v_2$ into~$\widehat{V_2}$.
    \item Put $v_2$ into~$V_2$ and $v_1, v_3, u$ into~$\widehat{V_2}$.
    \item Put $v_2,v_3$ into~$V_2$ and $v_1, u$ into~$\widehat{V_2}$.
\end{enumerate}
\end{brarule}

%\begin{toappendix}
\begin{lemma}
The branching is a complete case distinction. Moreover, it leads to a branching vector not worse than $(3 - \omega, 2 - \omega, 4,4)\,.$
\end{lemma}

\begin{pf}
Let $f: V \to \{0,1,2\}$ be a minimal rdf with $V_2 \subseteq V_2(f)$ and $\widehat{V_2}\cap V_2(f) = \emptyset$. The first case is equivalent to the first case in \autoref{BR_P1}. The second case is trivial. Assume $v_2\in V_2(f)$. Therefore, $u$ can not have the value~$2$. Otherwise, it has no private neighbor (except itself). In the last two cases, the measure decreases by 4 since $v_3$ has a fixed value, is dominated and $N[u]\cap \Active = \emptyset$.
\end{pf}
%\end{toappendix}

\begin{brarule}\label{BRP3}
Let $v_1$ be the leftmost vertex in $(\Active \cup \overline{V_2})$,  such that $ N[v_1]\cap \Active  = \{v_2\}$, with $N[v_2] \cap \overline{V_2} = \emptyset$ and $ N(v_2) \cap (\Active \cup \overline{V_2}) =\{v_1,v_3\}$. If there exists a $u$ that is leftmost in $\Active\setminus \{v_1,v_2,v_3\}$, with $\{v_3\} \subsetneq N(u)$, then branch as follows:

\begin{enumerate}
    \item Put $v_1$ in $V_2$ and $v_2,v_3$ in $\widehat{V_2}$.
    \item Put $v_1,v_2$ in $\widehat{V_2}$.
    \item Put $v_2$ in $V_2$ and $v_1, v_3$ in $\widehat{V_2}$.
    \item Put $v_2, v_3$ in $V_2$ and $v_1, N[u]\setminus \{ v_3 \}$ in $\widehat{V_2}$.
\end{enumerate}
\end{brarule}

%\begin{toappendix}
\begin{lemma}\label{lem:BRP3}
The branching is a complete case distinction. Moreover, it leads to a branching vector not worse than $(3 - \omega, 2 - \omega, 3, 5 - \omega)\,.$
\end{lemma}

\begin{pf}
We can assume there is no vertex $w\in N(v_3)$  with $N(w) = \{ v_3\}$. Otherwise, it would trigger  \autoref{BR_P2single}. Let $f: V \to \{0,1,2\}$ be a minimal rdf with $V_2 \subseteq V_2(f)$ and $\widehat{V_2}\cap V_2(f) = \emptyset$. The first two branchings are the same as in \autoref{BR_P2single}.
Assume that $f(v_1) \neq 2 = f(v_2)$. If $f(v_2) = 2\neq f(v_3)$, then vertex~$v_3$ is in $\widehat{V_2}\setminus \overline{V_2}$, since it is already dominated. This explains the third branching, including the decrease of the measure. 

The fourth branching needs most explanations, both concerning the completeness of the case distinction and the drop of the measure. 
Assume  that $f(v_1) \neq 2 = f(v_2) =f(v_3)$. Then, $v_3$ needs a private neighbor. Assume that this private  neighbor is a vertex $w\in N(v_3)\setminus \{ v_2, u\}$. We want to show that then $u$ is also a private neighbor. We distinguish two cases.\\
\underline{Case 1:} $I_u \subseteq I_{v_3}$.
This implies $u\notin V_2(f)$. We know $I_{v_2} \cap I_u = \emptyset$. If $u$ is not a private neighbor of $v_3$, then there exists a $y\in N(u)\cap V_2(f)$. Thus, $\emptyset \neq I_u\cap I_y \subseteq I_{v_3} \cap I_y$. Furthermore, we know $l_{v_3} < r_{v_2} < l_y$. Having $r_y<r_{v_3}$ would contradict \autoref{thm:property_min_rdf}. Therefore, we can assume $r_{v_3} < r_y$.  For all  $z\in (N(v_3) \cap (\Active \cup \overline{V_2}))\setminus (N[y] \cup \{v_2\})$, it holds that $r_{v_2} < l_z$ and that $r_z < l_y < r_u $. But this would contradict with the minimality of $r_u$.
Thus, $N[v_3]\setminus \{v_2\}\subseteq N[y]$, which then contradicts to the minimality of $f$. 
\\
\underline{Case 2:} $r_{v_3} < r_u$.
Assume there is a private neighbor $x \in (N(v_3)\setminus \{ u\}) \cap (\Active \cup \overline{V_2}).$ We want to show that $N(u)\subseteq N(x)$. Therefore, we assume $$l_u < l_x < r_{v_3} < r_u < r_x\,.$$ Let $z \in (N[u]\setminus N[x])\cap ( \Active \cup\overline{V_2})$. Thus, $I_u\cap I_z \subseteq I_u\setminus I_x = [l_u,l_x)$. Therefore, $r_z<l_x<r_u$ holds, which contradicts with the minimality of $u$.  This implies $N[u] \setminus (V_2\setminus \{ v_3 \})\subseteq N[x] \setminus (V_2\setminus \{ v_3 \}) =\emptyset$. 

Thus, we can conclude that $u$ is a private neighbor of $v_3$. Therefore,  $N[u] \setminus \{ v_3 \}\subseteq \widehat{V_2}$. Since $u$ has at least one neighbor that is not $v_3$, the measure loses $5 - \omega$.
\end{pf}
%\end{toappendix}

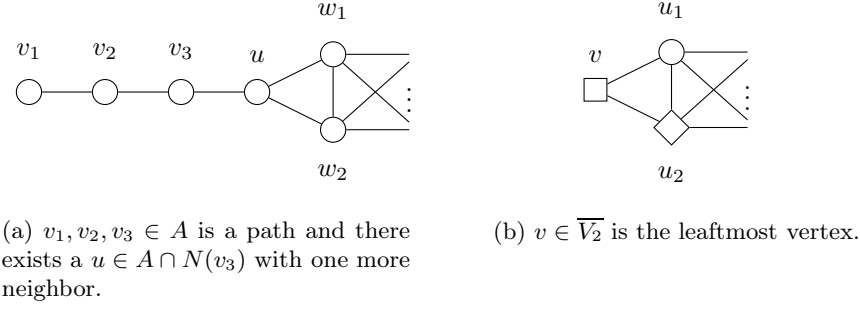
\begin{figure}[bt]
    \centering
\begin{subfigure}[t]{.44\textwidth}
    \centering
	\begin{tikzpicture}[transform shape]
			\tikzset{every node/.style={ fill = white,circle,minimum size=0.3cm}}
			\node[draw,label={above:$v_1$}] (v1) at (0,0) {};
			\node[draw,label={above:$v_2$}] (v2) at (1,0) {};
			\node[draw,label={above:$v_3$}] (v3) at (2,0) {};
			\node[draw,label={above:$u$}] (u) at (3,0) {};
			\node[draw,label={above:$w_1$}] (w1) at (4,0.5) {};
			\node[draw,label={below:$w_2$}] (w2) at (4,-0.5) {};
			\node[fill=none] at (5,0) {$\vdots$};
			\path (v1) edge[-] (v2);
			\path (v2) edge[-] (v3);
			\path (v3) edge[-] (u);
			\path (u) edge[-] (w1);
		    \path (u) edge[-] (w2);
		    \path (w1) edge[-] (w2);
			\path (w2) edge (5,0.4);
			\path (w1) edge (5,-0.4);
			\path (w2) edge (5,-0.5);
			\path (w1) edge (5,0.5);
        \end{tikzpicture}
    \subcaption{%Branching Rule~\ref{branching:dom-nbs-On}: 
    $v_1,v_2,v_3\in \Active$ is a path and there exists a $u\in \Active\cap N(v_3)$ with one more neighbor.}
    \label{fig:BRP3}
    \end{subfigure}
    \qquad
\begin{subfigure}[t]{.44\textwidth}
    \centering
    	
	\begin{tikzpicture}[transform shape]
		\tikzset{every node/.style={ fill = white,circle,minimum size=0.3cm}}
			\node[draw,rectangle,label={above:$v$}] (v) at (0,0) {};
			\node[draw,label={above:$u_1$}] (u1) at (1,0.5) {};
			\node[draw,diamond,label={below:$u_2$}] (u2) at (1,-0.5) {};
			\node[fill=none] at (2,0) {$\vdots$};
			\path (v) edge[-] (u1);
			\path (v) edge[-] (u2);
		    \path (u1) edge[-] (u2);
			\path (u2) edge (2,0.4);
			\path (u1) edge (2,-0.4);
			\path (u2) edge (2,-0.5);
			\path (u1) edge (2,0.5);

        \end{tikzpicture}

    \subcaption{%Branching Rule~\ref{branching:dom-nbs-On}: 
    $v\in \overline{V_2}$ is the leaftmost vertex.}
    \label{fig:BR_TildeV2}
\end{subfigure}

    \caption{Branching Rules~\ref{BRP3}  and~\ref{BR_TildeV2} }
\end{figure}
\begin{brarule}\label{BR_TildeV2}
Let $v$  be the leftmost vertex in $(\Active \cup \overline{V_2})$. If $v\in \widehat{V_2}$, branch as follows:

\begin{enumerate}
    \item For each $u\in N(v)\cap \Active$: $u$ in $V_2$ and $N(v)\setminus \{u\}$ into $\widehat{V_2}$.
    \item Put $N(v)$ in $\widehat{V_2}$.
\end{enumerate}
\end{brarule}

\begin{lemma}
The branching is a complete case distinction. Moreover, it leads  to a branching vector not worse than
$$(\underbrace{\omega + \vert N(v) \cap \Active \vert, \ldots, \omega + \vert N(v)\cap \Active\vert}_{\vert N(v) \cap \Active \vert \textrm{ many times}},\omega + (1 - \omega) \cdot \vert N(v)\cap \Active\vert)\, . $$
\end{lemma}

\begin{pf}
Let $f: V \to \{0,1,2\}$ be a minimal rdf with $V_2 \subseteq V_2(f)$ and $\widehat{V_2}\cap V_2(f) = \emptyset$. 
Assume there exists $x,y\in N(v) \cap \Active \cap  V_2(f) $ with $x\neq y$. Without loss of generality, $l_x < l_y < r_x < r_y$. By the minimality of $f$, there exists a $z\in N(x)\setminus N[y].$ This leads to $I_x\cap I_z \subseteq I_x \setminus I_y = [l_x,l_y)\subseteq [l_x,r_v)$. This contradicts to the minimality of $v$.

Therefore, at most one neighbor of $v$ is in $V_2(f)$. Furthermore, the minimality of $r_v$ implies $r_v\in I_u$ for $u\in \Active\cap N(v)$.  Since $\Active\cap N(v)$ is a clique, the measure is decreased by $\omega + \vert N(v)\cap \Active\vert$ if a vertex of $\Active\cap N(v)$ moves into $V_2$. If all vertices move into $\widehat{V_2}$, the measure loses $(1-\omega)\cdot \vert N(v)\cap \Active\vert$, because of the new vertices in $\widehat{V_2}$, and additionally $\omega$, since $v$ has no neighbors in~$\Active$ anymore.
\end{pf}

\begin{lemma}
For $n\in \mathbb{N}\setminus \{0\}$ and $\omega = 0.57$, the branching vector
$$(\underbrace{\omega + n, \ldots, \omega + n}_{n \text{ times}}, \omega + (1 - \omega) \cdot n )$$
has a branching number not greater than $\sqrt{3}$.
\end{lemma}

\begin{pf}
For the proof we will look at the corresponding polynomial at $\sqrt{3}$. 
We define 
$$g: \mathbb{R}^*_+ \to \mathbb{R}, n \mapsto \sqrt{3}^{n + \omega} - \sqrt{3}^{\omega \cdot n} - n.$$
The derivative is given by $g'(n)= \frac{\ln(3)}{2} \cdot \sqrt{3}^{n+\omega}- \frac{\ln(3)}{2} \cdot \omega \cdot \sqrt{3}^{\omega \cdot n}-1$ for $n\in \mathbb{R}^*_+$. Let $n\in [2,\infty)$. Then $g'(n)\geq  \frac{\ln(3)}{2} \cdot \sqrt{3}^{n} (1- \omega)-1$ holds. 
Furthermore,  $0\leq \frac{\ln(3)}{2} \cdot \sqrt{3}^{n} (1- \omega)-1$ holds for $n \geq 2 \cdot \log_3\left(\frac{2}{\ln(3) \cdot (1 - \omega)}\right).$ Therefore, $g$  is  increasing on $[2.6271, \infty)$.
Furthermore, $ 0.0011 \leq g(1)$, $0.2324\leq g(2)$ and $1.5483 \leq g(3)$ hold.
Thus, the branching number is smaller than or equal to $\sqrt{3}$.
\end{pf}

\begin{theorem}\label{mainInterval}
All minimal rdf of an interval graph of order~$n$ can be enumerated in time $\mathcal{O}^*(\sqrt{3}^n)$,  with polynomial delay and in polynomial space.
\end{theorem}
\begin{pf}
We get the branching numbers collected in \autoref{tab:Branching_interval-graphs}.
Therefore, we only need to show that the branching is a complete case distinction. Trivially, this holds if the vertices of $\overline{V_2}\cup\overline{V_1}$ can only appear in the neighborhood of the leftmost vertex of $\Active \cup \overline{V_1} \cup \overline{V_2}$. We will prove this by induction on the number of used branching rules. Since at the beginning each vertex is in $\Active$, this is settled. 

Assume that we are at some point of the algorithm and that each vertex of $\overline{V_2} \cup \overline{V_1}$ only appears in the neighborhood of the leftmost vertex of $\Active \cup \overline{V_1} \cup \overline{V_2}$. By induction, this also implies that each branching rule is sound at this moment $(*)$ (this is only important for the proof of \autoref{lem:BRDom}). If we add a vertex $v$ to $V_2$, each vertex $w$ with $r_w < r_v$ is either in $V_2$ or $\widehat{V_2}\setminus\overline{V_2}$ and the vertices with $l_w < r_v$ could also be in $\overline{V_1}$. After such a case, if the  vertex $u$  leftmost in $A\cup \overline{V_1} \cup \overline{V_2}$ is not in $\overline{V_1}$, $I_u \subseteq I_w$ holds for each $\overline{V_1}$. If we put no vertex in $V_2$, we put either the a part of the neighborhood of the leftmost vertex in $\widehat{V_2}$ (works analogously to $\overline{V_1}$ before or does not change anything). Therefore, vertices in $\overline{V_2}\cup\overline{V_1}$ can only appear in the neighborhood of the leftmost vertex of $\Active \cup \overline{V_1} \cup \overline{V_2}$.
\end{pf}

Notice that this result is optimal, as there are interval graphs that have $\sqrt{3}^n$ many minimal rdf, namely collections of paths.

\begin{table}[]
    \centering
   \begin{tabular}{c|c|c}
   rule  & branch. vector & branch. number\\ \hline
    \ref{BRDom} \& \ref{BRNotDominatable} & $(1,1 + \omega)$ & 1.7314\\
    %\ref{BRNotDominatable} & $(1 + \omega,1)$ & 1.7314\\
    \ref{BRP0} & $(3, 1 - \omega)$ & 1.6992\\
    \ref{BR_P2TildeV2} & $(2 + \omega, 2 + \omega, 2 -\omega)$ & 1.6829\\
    \ref{BR_P1} & $(4 - 2 \omega, 1 - \omega)$ & 1.7274\\
    \ref{BR_P2single} & $(3 - \omega, 2 - \omega, 4, 4)$ & 1.6877\\
    \ref{BRP3} & $(3 - \omega, 2 - \omega, 3, 5 - \omega)$ & 1.7315\\
    \ref{BR_TildeV2} & \small $(\underbrace{\omega + \vert N(v) \cap \Active \vert, \ldots, \omega + \vert N(v)\cap \Active\vert}_{\vert N(v) \cap \Active \vert \text{ many times}},\omega + (1 - \omega) \cdot \vert N(v)\cap \Active\vert)$ & \small $ \leq \sqrt{3}\leq 1.7321$ 
\end{tabular}
    \caption{Branching scenarios on interval graphs}
    \label{tab:Branching_interval-graphs}
\end{table}

\section{The Number of  Minimal Roman Dominating Functions on Forests}
\label{sec:min_rdf_forests}

Recall that a \emph{forest} is just a different name for an acyclic graph. 
The main part of this section is dedicated to the proof of the following theorem.

\begin{theorem}
A forest of order $n$ has at most $\sqrt{3}^n$ many minimal rdf. They can be also enumerated in this time,  with polynomial delay and in polynomial space. 
\end{theorem}

Notice that this result is optimal, as there are forests that have $\sqrt{3}^n$ many minimal rdf, namely collections of paths.
A similar optimality result was rather recently obtained by Günter Rote~\cite{Rot2019} for enumerating minimal dominating sets in forests: there are (at most) $\sqrt[13]{95}^n$ many of them in forests of order~$n$. 
We will prove our result by the construction of a branching algorithm. Let $G=(V,E)$ be a forest. As in \autoref{sec:min_rdf_interval}, our algorithm will partition the vertex set into $V_2$ and $\widehat{V_2}$.  Therefore, we will use the sets $\widehat{\Active}, V_2, \widehat{V_2}, \overline{V_2}$, where $\widehat{\Active}$ (represented by a~\tikz[fill lower half/.style={path picture={\fill[#1] (path picture bounding box.south west) rectangle (path picture bounding box.east);}}]{\draw (0,0) node[circle,fill lower half=black!40,scale=0.8, draw] {};} in the pictures) is the union of $\Active$ and $\overline{V_1}$ to save cases (the remaining sets have the same definitions).
Again, we have $\omega_1=1$ and $\omega_2=\omega=0.57$.
The measure is hence given by 
%$\mu = \mu (G,\widehat{\Active}, V_2, \widehat{V_2}, \overline{V_2}) = 
$\vert \widehat{\Active}\vert + \omega \;\vert \overline{V_2} \vert $. % (for some $\omega \in \left[ 0,1 \right] $). 
 Because of the construction, we can make use of \autoref{RRNoNeighbor} first. Then, the branching rules are applied, which are sometimes distinctively more complicated than in the case of interval graphs.
The set $L=L(G[\widehat{\Active}\cup \overline{V_2}])$ denotes the leaves (vertices of degree~$1$) of the forest $G[\widehat{\Active}\cup \overline{V_2}]$.

\begin{brarule}\label{BRLeafnot2}
Let $v\in \overline{V_2} \cap L$ with $u\in \widehat{\Active}\cap N(v)$. Branch as follows:

\begin{enumerate}
    \item Put $u$ in $V_2$.
    \item Put $u$ in $\widehat{V_2}$.
\end{enumerate}
\end{brarule}

%\begin{toappendix}
\begin{lemma}
The branching is a complete case distinction. Moreover, it leads  to a branching vector not worse than
$(1 + \omega, 1)\, . $
\end{lemma}

\begin{pf}
It is clear that this is a complete case distinction. In the first case, $u$ goes from $\widehat{\Active}$ to~$V_2$ and $v$ has a neighbor in $V_2$. Thus, $v\in \widehat{V_2}\setminus \overline{V_2}$. Hence, the measure decrases by $1+ \omega$.

In the second case, $u\in \widehat{V_2}$ and has no neighbor in $\widehat{\Active}$. Thus, $v$ has no neighbor in~$\widehat{\Active}$ anymore. This decreases the measure by~$1$.
\end{pf}
%\end{toappendix}

\begin{figure}[bt]
    \centering
    	
\begin{subfigure}[t]{.25\textwidth}
    \centering
    	
	\begin{tikzpicture}[transform shape]
			\tikzset{every node/.style={ fill = black,circle,minimum size=0.3cm},fill lower half/.style={path picture={\fill[#1] (path picture bounding box.south west) rectangle (path picture bounding box.east);}}}
			\node[draw,fill=white,rectangle,label={below:$v$}] (u) at (0,0) {};
			\node[draw,,fill=white,fill lower half=black!40,,label={below:$u$}] (v1) at (1,0) {};
			\path (u) edge[-] (v1);
			\path (v1) edge (2,0);

        \end{tikzpicture}

    \subcaption{%Branching Rule~\ref{branching:dom-nbs-On}: 
    $v\in\overline{V_2}$ is a leaf.}
    \label{fig:BRLeafnot2}
\end{subfigure}
\begin{subfigure}[t]{.25\textwidth}
    \centering
    	
	\begin{tikzpicture}[transform shape]
			\tikzset{every node/.style={ fill = black,circle,minimum size=0.3cm},fill lower half/.style={path picture={\fill[#1] (path picture bounding box.south west) rectangle (path picture bounding box.east);}}}
			\node[draw,,fill=white,fill lower half=black!40,,label={below:$v$}] (u) at (0,0) {};
			\node[draw,rectangle,fill=white,label={below:$u$}] (v1) at (1,0) {};
			\path (u) edge[-] (v1);
			\path (v1) edge (2,0);

        \end{tikzpicture}

    \subcaption{A leaf $v\in \widehat{\Active}$ with $N(v)=\{u\}\subseteq\overline{V_2}$.}
    \label{fig:BRLeafParentnot2}
    \end{subfigure}\ \ 
\begin{subfigure}[t]{.47\textwidth}
    \centering
    	
	\begin{tikzpicture}[transform shape]
			\tikzset{every node/.style={ fill = black,circle,minimum size=0.3cm},fill lower half/.style={path picture={\fill[#1] (path picture bounding box.south west) rectangle (path picture bounding box.east);}}}
			\node[draw,,fill=white,fill lower half=black!40,label={below:$w$}] (w) at (0,-0.5) {};
			\node[draw,,fill=white,fill lower half=black!40,label={below:$u$}] (u) at (0,0.5) {};
			\node[draw,,fill=white,fill lower half=black!40,label={below:$v$}] (v) at (1,0) {};
			\path (u) edge[-] (v);
			\path (w) edge[-] (v);
			\path (v) edge (2,0);

        \end{tikzpicture}

    \subcaption{$v$ has at least 2 pendant neighbors.}
    \label{fig:BRParentLeafs}
\end{subfigure}
    \caption{Illustrating Branching Rules~\ref{BRLeafnot2},~\ref{BRLeafParentnot2} and~\ref{BRParentLeafs}}
\end{figure}
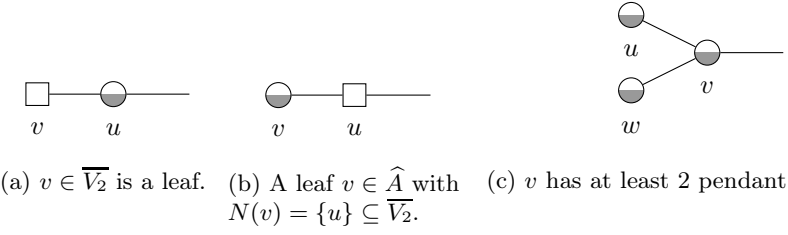

\begin{brarule}\label{BRLeafParentnot2}
Let $v\in \overline{V_2}$ such that there exists a $u\in \widehat{\Active}\cap N(v) \cap L$. Branch as follows:

\begin{enumerate}
    \item Put $u$ in $V_2$. 
    \item Put $u$ in $\widehat{V_2}$.
\end{enumerate}
\end{brarule}

\begin{lemma}
The branching is a complete case distinction. Moreover, it leads  to a branching vector not worse than $(1 + \omega, 1)\, . $
\end{lemma}

\begin{pf}
It is clear that this is a complete case distinction. In the first case, $u$ goes from $\widehat{\Active}$ to $V_2$ and $v$ has a neighbor in $V_2$. Thus, $v\in \widehat{V_2}\setminus \overline{V_2}$. Hence the measure decreases by $1+ \omega$.

In the second case, $u\in \widehat{V_2}$ (the measure decreases by $1- \omega$). Hence $u\in \widehat{V_2}\setminus \overline{V_2}$ (the measure decreases by $1$).
\end{pf}

\begin{brarule}\label{BRParentLeafs}
Let $v\in \widehat{\Active}$ with $\{u,w\}\subseteq N(v) \cap \widehat{\Active} \cap L$ and $u \neq w$. Branch as follows:

\begin{enumerate}
    \item Put $u$ in $V_2$ and $v, w$ in $\widehat{V_2}$. 
    \item Put $w$ in $V_2$ and $u,v$ in $\widehat{V_2}$. 
    \item Put $v$ in $V_2$ and $u ,w$ in $\widehat{V_2}$.
    \item Put $u,v,w$ in $\widehat{V_2}$.
\end{enumerate}
\end{brarule}

\begin{lemma}
The branching is a complete case distinction. Moreover, it leads to a branching vector not worse than 
$(3, 3, 3, 3 - \omega)\, . $
\end{lemma}

\begin{pf}
If $v\in V_2$ holds, \autoref{RRNoNeighbor} would put both $u$ and $w$ into $\widehat{V_2}\setminus \overline{V_2}$. Since $v,u,w$ were in $\widehat{\Active}$ before, this would reduce the measure by~$3$. 
Assume there exists a minimal rdf $f\in \{0,1,2\}^V$ with $V_2(f)\setminus \widehat{\Active} =V_2$ and $f(u)=f(w)=2$. This would contradict \autoref{thm:property_min_rdf}, since $v$ is the only possible private neighbor of $u$ and $w$.
Therefore, at most one of $u$ and $w$ can have the value~2. If one is in $V_2$, the other one is in $\widehat{V_2} \setminus \overline{V_2}$. Hence, in both cases the measure decreases by~3.
The last case ($u,v,w\in \widehat{V_2}$) decreases the measure by $3 - \omega$, since $u,w$ have no neighbor in $\widehat{\Active}$. 
\end{pf}

\noindent
From now on, we can assume that there exists no $v\in\widehat{\Active}\cup \overline{V_2}$ with $\vert N(v)\cap L \cap (\widehat{\Active}\cup \overline{V_2})\vert> 1$.

\begin{brarule}\label{BRP2vnot2}
Let $v\in \overline{V_2}$ with $ u \in N(v) \cap \widehat{\Active}$, $\{ w \}= N(u) \cap L$ and $\vert N(u)\cap (\widehat{\Active}\cup \overline{V_2}) \vert = 2$. Branch as follows:

\begin{enumerate}
    \item Put $u$ in $V_2$ and $w$ in $\widehat{V_2}$. 
    \item Put $w$ in $V_2$ and $u$ in $\widehat{V_2}$. 
    \item Put $u,w$ in $\widehat{V_2}$.
\end{enumerate}
\end{brarule}

\begin{lemma}
The branching is a complete case distinction. Moreover, it leads  to a branching vector not worse than
$(2 + \omega, 2, 2)\, . $
\end{lemma}

\begin{pf}
By  \autoref{RRNoNeighbor}, we know: If $u\in V_2$, $w$ has to be in $\widehat{V_2}\setminus \overline{V_2}$. Therefore, this is a complete case distinction.
If $u\in V_2$, then $w,v\in \widehat{V_2}\setminus\overline{V_2}$  and the measure decreases by~$2+\omega$.
In the second case, the measure is reduced by~$2$, since $u$ is dominated by $w$.
As $N[\{u,w\}]\subseteq \{v,u,w\}$, $u,w\in \widehat{V_2}\setminus \overline{V_2}$,  the measure decreases by~2 also in the third case.
\end{pf}

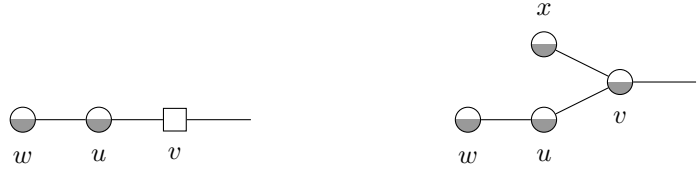
\begin{figure}[bt]
    \centering
    	
\begin{subfigure}[t]{.43\textwidth}
    \centering
    	
	\begin{tikzpicture}[transform shape]
			\tikzset{every node/.style={ fill = white,circle,minimum size=0.3cm},fill lower half/.style={path picture={\fill[#1] (path picture bounding box.south west) rectangle (path picture bounding box.east);}}}
			\node[draw,rectangle,label={below:$v$}] (v) at (1,0) {};
			\node[draw,fill lower half=black!40,label={below:$u$}] (u) at (0,0) {};
			\node[draw,fill lower half=black!40,label={below:$w$}] (w) at (-1,0) {};
			\path (u) edge[-] (v);
			\path (w) edge[-] (u);
			\path (v) edge (2,0);

        \end{tikzpicture}

    \subcaption{$w,u\in \widehat{\Active}$ and $v\in \overline{V_2}$ build a path where $w$ is a leaf.}
    \label{fig:BRP2not2}
\end{subfigure}
\begin{subfigure}[t]{.51\textwidth}
    \centering
    	
	\begin{tikzpicture}[transform shape]
			\tikzset{every node/.style={ fill = white,circle,minimum size=0.3cm},fill lower half/.style={path picture={\fill[#1] (path picture bounding box.south west) rectangle (path picture bounding box.east);}}}
			\node[draw,fill lower half=black!40,label={below:$v$}] (v) at (1,0) {};
			\node[draw,fill lower half=black!40,label={below:$u$}] (u) at (0,-0.5) {};
			\node[draw,fill lower half=black!40,label={above:$x$}] (x) at (0,0.5) {};
			\node[draw,fill lower half=black!40,label={below:$w$}] (w) at (-1,-0.5) {};
			\path (u) edge[-] (v);
			\path (w) edge[-] (u);
			\path (x) edge[-] (v);
			\path (v) edge (2,0);
        \end{tikzpicture}

    \subcaption{$u,w,v\in \widehat{\Active}$ are a path where $w$ is a leaf and $v$ has a leaf as neighbor in $\widehat{\Active}$.}
    \label{fig:BRP2ParentLeaf}
    \end{subfigure}

    \caption{Picturing Branching Rules~\ref{BRP2vnot2} and~\ref{BRP2ParentLeaf}}
\end{figure}

\begin{brarule}\label{BRP2ParentLeaf}
Let $v\in \widehat{\Active}$ with $ u \in N(v) \cap \widehat{\Active}$, $\{ w \}= N(u) \cap L\cap \widehat{\Active}$, $\vert N(u)\cap (\widehat{\Active}\cup \overline{V_2}) \vert = 2$ and $\{x\}=L \cap N(v)\cap \widehat{\Active}$. Branch as follows: 

\begin{enumerate}
    \item Put $v$ in $V_2$ and $w,x$ in $\widehat{V_2}$. 
    \item Put $x$ in $V_2$ and $v,u$ in $\widehat{V_2}$. 
    \item Put $v,x$ in $\widehat{V_2}$.
\end{enumerate}
\end{brarule}

\begin{lemma}
The branching is a complete case distinction. Moreover, it leads  to a branching vector not worse than
$(3 - \omega, 3 - \omega, 2 - \omega)\, . $
\end{lemma}

\begin{pf}
By  \autoref{RRNoNeighbor}, we know that, if $v\in V_2$ holds, then $x,w$ have to be in $\widehat{V_2}$. In this case, $x$ is already dominated. Thus, the measure is decreased by $3- \omega$.
Assume there exists a minimal rdf $f\in \{0,1,2\}^V$ with $V_2(f)\setminus \widehat{\Active} =V_2$ and $f(u)=f(x)=2$. This would contradict \autoref{thm:property_min_rdf}, since $v$ is the only possible private neighbor of~$x$.
In the third case, the measure is reduced by $2-\omega$, since vertex~$x$ has no longer any neighbor in~$\widehat{\Active}$.
\end{pf}

\begin{figure}[bt]
    \centering
    	
\begin{subfigure}[b]{.46\textwidth}
    \centering
    	
	\begin{tikzpicture}[transform shape]
			\tikzset{every node/.style={ fill = white,circle,minimum size=0.3cm},fill lower half/.style={path picture={\fill[#1] (path picture bounding box.south west) rectangle (path picture bounding box.east);}}}
			\node[draw,fill lower half=black!40,label={below:$v$}] (v) at (1,0) {};
			\node[draw,fill lower half=black!40,label={below:$u_1$}] (u1) at (0,-0.5) {};
			\node[draw,fill lower half=black!40,label={above:$u_2$}] (u2) at (0,0.5) {};
			\node[draw,fill lower half=black!40,label={below:$w_1$}] (w1) at (-1,-0.5) {};
			\node[draw,fill lower half=black!40,label={above:$w_2$}] (w2) at (-1,0.5) {};
			\path (u1) edge[-] (v);
			\path (u2) edge[-] (v);
			\path (w1) edge[-] (u1);
			\path (w2) edge[-] (u2);
			\path (v) edge (2,0);

        \end{tikzpicture}

    \subcaption{$w_i,u_i,v\in \widehat{\Active}$ are paths; $w_1,w_2$ are leaves.}
    \label{fig:BR2P2}
\end{subfigure}\ \ 
\begin{subfigure}[b]{.5\textwidth}
    \centering
    	
	\begin{tikzpicture}[transform shape]
			\tikzset{every node/.style={ fill = white,circle,minimum size=0.3cm},fill lower half/.style={path picture={\fill[#1] (path picture bounding box.south west) rectangle (path picture bounding box.east);}}}
			\node[draw,rectangle,label={below:$v$}] (v) at (1,0) {};
			\node[draw,fill lower half=black!40,label={below:$u$}] (u) at (0,0) {};
			\node[draw,fill lower half=black!40,label={below:$x$}] (x) at (-2,0) {};
			\node[draw,fill lower half=black!40,label={below:$w$}] (w) at (-1,0) {};
			\path (u) edge[-] (v);
			\path (w) edge[-] (u);
			\path (x) edge[-] (w);
			\path (v) edge (2,0);
        \end{tikzpicture}

    \subcaption{$x,w,u\in \widehat{\Active}$, $v\in \overline{V_2}$ form a path and $x$ is a leaf.}
    \label{fig:BRP3vnot2}
    \end{subfigure}

    \caption{Illustrating Branching Rules~\ref{BR2P2} and~\ref{BRP3vnot2}}
\end{figure}
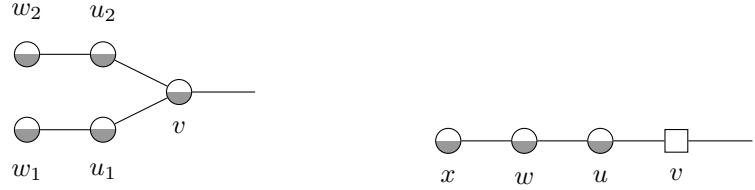

Now, we present a seemingly complicated rule, but it is  a rather complete branching when two paths meet in a vertex, see \autoref{fig:BR2P2}.

\begin{brarule}\label{BR2P2}
Let $v\in \widehat{\Active}$ with $ u_1,u_2 \in N(v) \cap \widehat{\Active}$, $\{ w_1 \}= N(u_1) \cap L\cap \widehat{\Active}$ and $\{ w_2 \}= N(u_2) \cap L\cap \widehat{\Active}$. Branch as follows:

\begin{enumerate}
    \item Put $v,u_1,u_2$ in $V_2$ and $w_1,w_2$ in $\widehat{V_2}$.
    \item Put $v,u_1$ in $V_2$ and $u_2,w_1,w_2$ in $\widehat{V_2}$.
    \item Put $v,u_2$ in $V_2$ and $u_1,w_1,w_2$ in $\widehat{V_2}$.
    \item Put $v$ in $V_2$ and $u_1,u_2,w_1,w_2$ in $\widehat{V_2}$.
    \item Put $u_1$ in $V_2$ and $v,w_1$ in $\widehat{V_2}$.
    \item Put $u_2,w_1$ in $V_2$ and $v,u_1,w_2$ in $\widehat{V_2}$.
    \item Put $w_2,w_1$ in $V_2$ and $v,u_1,u_2$ in $\widehat{V_2}$.
    \item Put $w_1$ in $V_2$ and $v,u_1,u_2,w_2$ in $\widehat{V_2}$.
    \item Put $u_2$ in $V_2$ and $v,u_1,w_1,w_2$ in $\widehat{V_2}$.
    \item Put $w_2$ in $V_2$ and $v,u_1,u_2,w_1$ in $\widehat{V_2}$.   
    \item Put $v,u_1,u_2,w_1,w_2$ in $\widehat{V_2}$.
\end{enumerate}
\end{brarule}

\begin{lemma}
The branching is a complete case distinction. Moreover, it leads to a branching vector not worse than
$(5,5,5,5,3,5,5-\omega,5-\omega,5,5-\omega,5-\omega)\, . $
\end{lemma}

\begin{pf}
By \autoref{RRNoNeighbor}, $w_1, w_2$ have to be in $\widehat{V_2}$, if $v \in V_2$ holds. Therefore, the first four cases are a complete case distinction for $v\in V_2$. In all these cases, no vertex is in $\overline{V_2}$. Thus, the measure is decreased by 5 in each case.

For the remaining 7 cases, we assume $v\in \widehat{V_2}$. This implies that $v,u_1,w_1$ fulfills the requirement of \autoref{BRP2vnot2}. This implies the branching vector $(5, 5, 5, 5, 3, 3 - \omega, 3 - \omega)$. For the last 2 cases of this branching vector, $v\in \overline{V_2}$ holds. Hence, we can use again  \autoref{BRP2vnot2}. This implies the branching vector $(5,5,5,5,3,5,5-\omega,5-\omega,5,5-\omega,5-\omega)\, . $ 
\end{pf}

\begin{brarule}\label{BRP3vnot2}
Let $v\in \overline{V_2}$ with $ u \in N(v) \cap \widehat{\Active}$, $\{ w,v \}= N(u)$, $\{ x \}= N(w) \cap L$ and $\vert N(w)\cap (\widehat{\Active}\cup \overline{V_2}) \vert = 2$. Branch as follows:

\begin{enumerate}
    \item Put $u$ in $V_2$ and $x$ in $\widehat{V_2}$. 
    \item Put $w$ in $V_2$ and $u,x$ in $\widehat{V_2}$. 
    \item Put $u,w$ in $\widehat{V_2}$.
\end{enumerate}
\end{brarule}

\begin{lemma}
The branching is a complete case distinction. Moreover, it leads to a branching vector not worse than
$(2, 3, 2 - \omega)\, . $
\end{lemma}

\begin{pf}
By \autoref{RRNoNeighbor}, $x$ has to be in $\widehat{V_2}$, if $u$ or $w$ are in $V_2$ holds. With this information, this becomes a asymmetric branching on $u,w$. Therefore this is a complete case distinction. In the first case, $v$ is dominated by $u$. This implies that the measure decreases by 2.  

In the second case, $u$ and $x$ are dominated by $w$. Hence, $\mu$ is reduced by 3. For the last case, $u$ is in $\widehat{V_2} \setminus \overline{V_2}$, as it has no neighbor in $\widehat{\Active}$. This reduces the measure by $2 - \omega$.
\end{pf}

\begin{brarule}\label{BRP3ParentLeaf}
Let $v\in \widehat{\Active}$ with $y\in N(v)\cap L$, $ u \in N(v) \cap \widehat{\Active}$, $\{ w,v \}= N(u)$, $\{ x \}= N(w) \cap L$ and $\vert N(w)\cap (\widehat{\Active}\cup \overline{V_2}) \vert = 2$. Branch as follows:

\begin{enumerate}
    \item Put $v$ in $V_2$ and $y$ in $\widehat{V_2}$. 
    \item Put $u$ in $V_2$ and $v,x$ in $\widehat{V_2}$.
    \item Put $w$ in $V_2$ and $v,u,x$ in $\widehat{V_2}$.
    \item Put $x$ in $V_2$ and $v,u,w$ in $\widehat{V_2}$. 
    \item Put $v,u,w,x$ in $\widehat{V_2}$.
\end{enumerate}
\end{brarule}

\begin{lemma}
The branching is a complete case distinction. Moreover, it leads to a branching vector not worse than
$(2,4-\omega,4-\omega,4-\omega,4-\omega)\, . $
\end{lemma}

\begin{pf}
By \autoref{RRNoNeighbor}, if $v\in V_2$, then~$y$ can not have the value~2. Furthermore,~$y$ is dominated and in $\widehat{V_2}\setminus \overline{V_2}$. This decreases the measure by~2. For the remaining proof, we assume $v\in\widehat{V_2}$. If $u\in V_2$, then we can use \autoref{RRNoNeighbor} on~$x$ and~$y$. In this case $y$ has no neighbor in $\widehat{\Active}$ and is in $\widehat{V_2}\setminus \overline{V_2}$. This reduces the measure by $4-\omega$. For $v,u\in \widehat{V_2}$, we can use \autoref{BRP2vnot2}. In this case, the branching vector better than in the other uses of \autoref{BRP2vnot2}, since~$u$ has no neighbor in $\widehat{\Active}$ after using this branch. Thus, in each of the three cases, the measure is decreasing by $4-\omega$. 
\end{pf}

 \begin{figure}[bt]
    \centering
    	
\begin{subfigure}[t]{.43\textwidth}
    \centering
    	
	\begin{tikzpicture}[transform shape]
			\tikzset{every node/.style={circle,minimum size=0.3cm}}
		\tikzset{every node/.style={ fill = white,circle,minimum size=0.3cm},fill lower half/.style={path picture={\fill[#1] (path picture bounding box.south west) rectangle (path picture bounding box.east);}}}
			\node[draw,fill lower half=black!40,label={below:$v$}] (v) at (1,0) {};
			\node[draw,fill lower half=black!40,label={above:$y$}] (y) at (0,0.5) {};
			\node[draw,fill lower half=black!40,label={below:$u$}] (u) at (0,-0.5) {};
			\node[draw,fill lower half=black!40,label={below:$x$}] (x) at (-2,-0.5) {};
			\node[draw,fill lower half=black!40,label={below:$w$}] (w) at (-1,-0.5) {};
			\path (y) edge[-] (v);
			\path (u) edge[-] (v);
			\path (w) edge[-] (u);
			\path (x) edge[-] (w);
			\path (v) edge (2,0);

        \end{tikzpicture}

    \subcaption{$x,w,u,v\in \widehat{\Active}$ induce a path and $v$ has a neighbor of degree one.}
    \label{fig:BRP3ParentLeaf}
\end{subfigure}\ \ 
\begin{subfigure}[t]{.46\textwidth}
    \centering
    	
	\begin{tikzpicture}[transform shape]
			\tikzset{every node/.style={ fill = white,circle,minimum size=0.3cm},fill lower half/.style={path picture={\fill[#1] (path picture bounding box.south west) rectangle (path picture bounding box.east);}}}
			\node[draw,fill lower half=black!40,label={below:$v$}] (v) at (1,0) {};
			\node[draw,fill lower half=black!40,label={above:$y$}] (y) at (0,0.5) {};
			\node[draw,fill lower half=black!40,label={above:$z$}] (z) at (-1,0.5) {};
			\node[draw,fill lower half=black!40,label={below:$u$}] (u) at (0,-0.5) {};
			\node[draw,fill lower half=black!40,label={below:$x$}] (x) at (-2,-0.5) {};
			\node[draw,fill lower half=black!40,label={below:$w$}] (w) at (-1,-0.5) {};
			\path (y) edge[-] (v);
			\path (y) edge[-] (z);
			\path (u) edge[-] (v);
			\path (w) edge[-] (u);
			\path (x) edge[-] (w);
			\path (v) edge (2,0);
        \end{tikzpicture}

    \subcaption{$x,w,u,v\in \widehat{\Active}$ and $z,y,v\in \Active$ are paths, where $z,x$ are leaves.}
    \label{fig:BRP3P2}
    \end{subfigure}

    \caption{Branching Rules~\ref{BRP3ParentLeaf} and~\ref{BRP3P2}}
\end{figure}
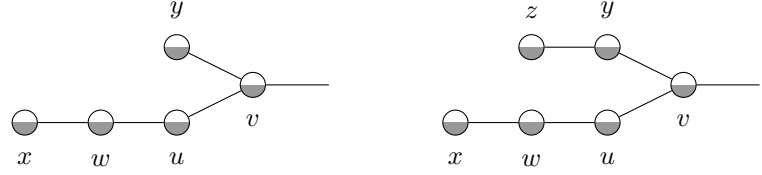
\begin{brarule}\label{BRP3P2}
Let $v\in \widehat{\Active}$ with $ u,y \in N(v) \cap \widehat{\Active}$, $\{ w,v \}= N(u)$, $\{ x \}= N(w) \cap L$, $z\in N(y)\cap L$ and $\vert N(w)\cap (\widehat{\Active} \cup \overline{V_2}) \vert = 2 = \vert N(y)\cap (\widehat{\Active} \cup \overline{V_2}) \vert$. Branch as follows:

\begin{enumerate}
    \item Put $v,u$ in $V_2$ and $w,x,z$ in $\widehat{V_2}$. 
    \item Put $v,y$ in $V_2$ and $u,z$ in $\widehat{V_2}$.
    \item Put $v$ in $V_2$ and $u,y,z$ in $\widehat{V_2}$.
    \item Put $u,w$ in $V_2$ and $v,x,y$ in $\widehat{V_2}$.
    \item Put $u$ in $V_2$ and $v,w,x$ in $\widehat{V_2}$.
    \item Put $w$ in $V_2$ and $v,u,x$ in $\widehat{V_2}$.
    \item Put $x$ in $V_2$ and $v,u,w$ in $\widehat{V_2}$. 
    \item Put $v,u,w,x$ in $\widehat{V_2}$.
\end{enumerate}
\end{brarule}

\begin{lemma}
The branching is a complete case distinction. Moreover, it leads to a branching vector not worse than
$(5 - \omega, 4, 4, 5 - \omega, 4, 4 - \omega, 4 - \omega,4 - \omega)\, . $
\end{lemma}

\begin{pf}
Let $f\in \{0,1,2\}^V$ be a minimal rdf with $V_2(f)\cap(V_2 \cup \widehat{V_2})=V_2$. Assume $f(v)=f(u)=2$. By \autoref{thm:property_min_rdf}, $w$ has to be the private neighbor of $u$ and to neither $f(w)$ nor $f(x)$, we can assign~2. \autoref{RRNoNeighbor} implies $f(z)\neq 2$ for each minimal rdf with $V_2(f)\cap(V_2 \cup \widehat{V_2})=V_2$ and $f(v)=2$. Therefore, the first three branches are a complete case distinction for $v\in V_2$.  
 For the remaining branches, we assume $f(v)\neq 2$. If $f(u)=f(w)=2$ holds, $x$ has to be a private neighbor of $w$ and $v$ has to be a private neighbor of $u$. This implies $x,y,v\in \widehat{V_2}(f)$. By \autoref{RRNoNeighbor}, the branches 4 and 5 are a complete case distinction for $v\in \widehat{V_2}$ and $u\in V_2$. The remaining 3 branches are induced by $v,u\in\widehat{V_2}$ and \autoref{BRP2vnot2}. After this branch, $u$ has no neighbor in $\widehat{\Active}$ anymore. Hence, in all of these 3 branches, the measure is decreased by $4-\omega$. 
 
In the first case, each vertex gets into $V_2 \cup \left( \widehat{V_2} \setminus \overline{V_2} \right)$, except~$z$. Therefore, the measure is decreased by $5-\omega$.
In the second case, each new vertex in $\widehat{V_2}$ is dominated. Thus, the measure is reduced by 4. 
In the third branch, $z$ has no neighbor in $\widehat{\Active}$. This implies that $\mu$ is decreased by~4.
In the fourth case each new vertex in $\widehat{V_2}$ is dominated except $y$. This implies that the measure is decreased by $5-\omega$. In the fifth branch, the measure is reduced by 4, as all new vertices in~$\widehat{V_2}$ are dominated except $x$, which has no neighbor in $\widehat{\Active}$ anymore.  
\end{pf}

\begin{brarule}\label{BR2P3}
Let $v\in \widehat{\Active}$ with $ u_1,u_2 \in N(v) \cap \widehat{\Active}$, $\{ w_i,v_i \}= N(u_i)$, $\{ x_i \}= N(w_i) \cap L$,  and $\vert N(w_i)\cap (\widehat{\Active}\cup \overline{V_2}) \vert = 2$ for $i\in \{1,2\}$. Branch as follows:

\begin{enumerate}
    \item Put $u_1,u_2$ in $V_2$ and $w_1,w_2,x_1,x_2$ in $\widehat{V_2}$.
    \item Put $u_1,w_1$ in $V_2$ and $v,u_2,x_1$ in $\widehat{V_2}$.
    \item Put $u_1$ in $V_2$ and $u_2,w_1,x_1$ in $\widehat{V_2}$.
    \item Put $w_1$ in $V_2$ and $u_1,x_1,$ in $\widehat{V_2}$.
    \item Put $x_1$ in $V_2$ and $u_1,w_1,$ in $\widehat{V_2}$. 
    \item Put $u_1,w_1,x_1$ in $\widehat{V_2}$. 
\end{enumerate}
\end{brarule}

\begin{lemma}
The branching is a complete case distinction. Moreover, it leads to a branching vector not worse than
$(6,5-\omega,4-\omega,3,3-\omega,3-\omega)\, . $
\end{lemma}

\begin{pf}
Let $f\in \{0,1,2\}^V$ be a minimal rdf with $V_2(f)\cap(V_2 \cup \widehat{V_2})=V_2$. If $f(u_1)=f(u_2)=2$ holds, $u_1$ and $u_2$ need a private neighbor. The only possibilties are $w_1,w_2$. Therefore, for each $t\in N[\{w_1,w_2\}]\setminus\{u_1,u_2\}=\{w_1,w_2,x_1,x_2\}$, $f(t)\neq 2$ holds. By construction, $w_1,w_2,x_1,x_2$ would be in $\widehat{V_2}\setminus\overline{V_2}$. This reduces the measure by $6$.

If $f(u_1)=f(w_1)=2\neq f(u_2)$ holds, $v$ is the only possible private neighbor of $u_1$ and $x_1$ the only one for~$w_1$. This implies $f(v)=f(x_1)=0$. Thus, the measure is decreased by $5-\omega$. 

Assume $f(u_1)=2\neq f(u_2)$ and $f(w_1)=0$ (this holds since it is a neighbor of $u_1$). In this case, $f(x_1)\neq 2$, as $x_1$ has no private neighbor. Furthermore, all neighbors of $x_1$ are now decided. Hence, the measure is decreased by $4-\omega$. 

For the last three cases, we assume $u_1\in \widehat{V_2}$ and use \autoref{BRP2vnot2}.
\end{pf}

 \begin{figure}[bt]
    \centering
    	
\begin{subfigure}[b]{.43\textwidth}
    \centering
    	
	\begin{tikzpicture}[transform shape]
			\tikzset{every node/.style={ fill = white,circle,minimum size=0.3cm},fill lower half/.style={path picture={\fill[#1] (path picture bounding box.south west) rectangle (path picture bounding box.east);}}}
			\node[draw,fill lower half=black!40,label={below:$v$}] (v) at (1,0) {};
			\node[draw,fill lower half=black!40,label={above:$u_2$}] (u2) at (0,0.5) {};
			\node[draw,fill lower half=black!40,label={above:$w_2$}] (w2) at (-1,0.5) {};
			\node[draw,fill lower half=black!40,label={below:$u_1$}] (u1) at (0,-0.5) {};
			\node[draw,fill lower half=black!40,label={below:$x_1$}] (x1) at (-2,-0.5) {};
			\node[draw,fill lower half=black!40,label={above:$x_2$}] (x2) at (-2,0.5) {};
			\node[draw,fill lower half=black!40,label={below:$w_2$}] (w1) at (-1,-0.5) {};
			\path (u1) edge[-] (v);
			\path (u2) edge[-] (v);
			\path (w1) edge[-] (u1);
			\path (w2) edge[-] (u2);
			\path (x1) edge[-] (w1);
			\path (x2) edge[-] (w2);
			\path (v) edge (2,0);

        \end{tikzpicture}

    \subcaption{$x_i,w_i,u_i,v\in \widehat{\Active}$ are paths, where $x_1,x_2$ are leaves.}
    \label{fig:BR2P3}
\end{subfigure}\ \ 
\begin{subfigure}[b]{.48\textwidth}
    \centering
    	
	\begin{tikzpicture}[transform shape]
			\tikzset{every node/.style={ fill = white,circle,minimum size=0.3cm},fill lower half/.style={path picture={\fill[#1] (path picture bounding box.south west) rectangle (path picture bounding box.east);}}}
			\node[draw,rectangle,label={below:$v$}] (v) at (1,0) {};
			\node[draw,fill lower half=black!40,label={below:$u$}] (u) at (0,0) {};
			\node[draw,fill lower half=black!40,label={below:$w$}] (w) at (-1,0) {};
			\node[draw,fill lower half=black!40,label={below:$x$}] (x) at (-2,0) {};
			\node[draw,fill lower half=black!40,label={below:$y$}] (y) at (-3,0) {};
			\path (y) edge[-] (x);
			\path (u) edge[-] (v);
			\path (w) edge[-] (u);
			\path (x) edge[-] (w);
			\path (v) edge (2,0);
        \end{tikzpicture}

    \subcaption{$y,x,w,u\in \widehat{\Active}$, $v\in \overline{V_2}$ is a path and $y$ is a leaf.}
    \label{fig:BRP4not}
    \end{subfigure}

    \caption{Branching Rules~\ref{BR2P3} and~\ref{BRP4not2}}
\end{figure}
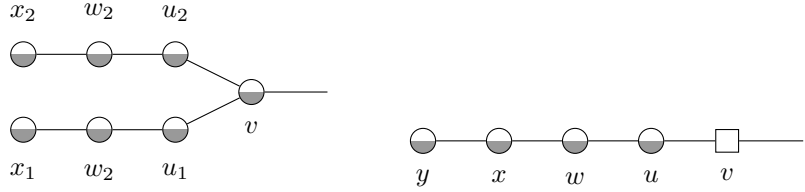

\begin{brarule}\label{BRP4not2}
Let $v\in \overline{V_2}$ with $ u \in N(v) \cap \widehat{\Active}$, $\{ w,v \}= N(u)$, $\{ x,u \}= N(w)$, $\{y\}=N(x)\cap L$ and $\vert N(x)\cap (\widehat{\Active}\cup \overline{V_2}) \vert = 2$. Branch as follows:

\begin{enumerate}
    \item Put $u,w$ in $V_2$ and $x,y$ in $\widehat{V_2}$. 
    \item Put $u$ in $V_2$ and $w$ in $\widehat{V_2}$. 
    \item Put $w,x$ in $V_2$ and $u,y$ in $\widehat{V_2}$.
    \item Put $w$ in $V_2$ and $u,x,y$ in $\widehat{V_2}$.
    \item Put $x$ in $V_2$ and $u,w,y$ in $\widehat{V_2}$.
    \item Put $u,w,x$ in $\widehat{V_2}$.
\end{enumerate}
\end{brarule}

\begin{lemma}
The branching is a complete case distinction. Moreover, it leads to a branching vector not worse than
$(4+\omega, 2+\omega,4,4,4,3-\omega)\, . $
\end{lemma}

\begin{pf}
Let $f\in \{0,1,2\}^V$ be a minimal rdf with $V_2(f)\cap(V_2 \cup \widehat{V_2})=V_2$. If $f(u)=f(w)=2$ holds, $x$ has to be a private neighbor of $w$. Therefore, $f(x)\neq 2$ and $f(y)\neq 2$ hold. This reduces the measure by $4+\omega$, since $y$ has no neighbor in $\widehat{\Active}$ and $v,x$ are dominated.

For $f(u)=2\neq f(w)$, the measure decreases by $2+\omega$, as $v,w$ are dominated by $u$.

For the remaining proof, we assume $f(u)\neq 2$. $f(w) = 2$ implies $f(y)\neq2$, as $y$ would not have a private neighbor. Thus, the measure is decreased by~$4$ in the branches~3 and~4, as $y$ has no neighbor in $\widehat{\Active}$. For $f(u)\neq 2$, $f(w)\neq 2$ and $f(x)=2$, $f(y)=0$ must hold. Since $u,w,y$ have no neighbor in $\widehat{\Active}$, $\mu$ is reduced by~4.
In the remaining case $u,w$ do not have any neighbor in $\widehat{\Active}$. Therefore, the measure is decreased by $3-\omega$. 
\end{pf}

\begin{brarule}\label{BRP3Tree}
Let $v\in \widehat{\Active}$ with $ u \in N(v) \cap \widehat{\Active}$, $\{ w,v \}= N(u)$, $\{ x,u \}= N(w)$, $\{y\}=N(x)\cap L$ and $\vert N(x)\cap (\widehat{\Active}\cup \overline{V_2}) \vert = 2$. Branch as follows:

\begin{enumerate}
    \item Put $y$ in $V_2$ and $x,w$ in $\widehat{V_2}$.
    \item Put $y,x$ in $\widehat{V_2}$.
    \item Put $x$ in $V_2$ and $y, w$ in $\widehat{V_2}$.
    \item Put $x,w$ in $V_2$ and $y,u,v$ in $\widehat{V_2}$.
\end{enumerate}
\end{brarule}

\begin{lemma}
The branching is a complete case distinction. Moreover, it leads to a branching vector not worse than $(3 - \omega, 2 - \omega, 3, 5 - \omega)\,.$
\end{lemma}
The proof is a special case of the proof of  \autoref{lem:BRP3} and hence omitted.  

\begin{table}\centering
\begin{tabular}{c|c|c}
   Rule  & branching vector & branching number\\ \hline
   \ref{BRLeafnot2} \& \ref{BRLeafParentnot2}      & $(1+\omega,1)$                                            & 1.7314\\
   \ref{BRParentLeafs}    & $(3,3,3,3-\omega)$                                        & 1.6288\\
   \ref{BRP2vnot2}        & $(2,2,2+\omega)$                                          & 1.6582\\
   \ref{BRP2ParentLeaf}   & $(3-\omega,3-\omega,2-\omega)$                            & 1.7164\\
   \ref{BR2P2}            & $(5,5,5,5,3,5,5-\omega,5-\omega,5,5-\omega,5-\omega)$     & 1.7029\\
   \ref{BRP3vnot2}        & $(2,3,2-\omega)$                                          & 1.7156\\
   \ref{BRP3ParentLeaf}   & $(2,4-\omega,4-\omega,4-\omega,4-\omega)$                 & 1.6966\\
   \ref{BRP3P2}           & $(5- \omega, 4,4,5-\omega, 4,4-\omega,4-\omega,4-\omega)$ & 1.7158\\
   \ref{BR2P3}            & $(6,5-\omega,4-\omega,3,3-\omega,3-\omega)$               & 1.7296\\
   \ref{BRP4not2}         & $(4+\omega,2+\omega,4,4,4,3-\omega)$                      & 1.7275\\
   \ref{BRP3Tree}             & $(3 - \omega, 2 - \omega, 3, 5 - \omega)$                 & 1.7315
    
\end{tabular}

\caption{\label{tab:forest-branchings}Branching vectors and numbers for enumerating minimal rdf in forests.}
\end{table}

\section{Enumerating Minimal RDF
in Chordal Graphs}\label{SecChordalEnum}

Recall that a graph is \emph{chordal} if the only induced cycles it might contain have length three. 
In this section, we are going to prove the following result:

\begin{theorem}
All minimal Roman dominating functions of a chordal graph of order~$n$ can be enumerated with polynomial delay and in polynomial space in time $\Oh(\RomanChordalUpperbound^n)$.
\end{theorem}

\noindent 
We are following the general approach sketched in \autoref{sec:general}.
For our branching scenario, we adopt as measure $\mu = |\Active| + \omega_1\, |\overline{V_1}| + \omega_2\, | \overline{V_2}|$.
To obtain our result, we set $\omega_1=0.710134$ and $\omega_2=0.434799$.

\medskip
\noindent
Initially all vertices are in $\Active$. Each branching rule assumes the preceding rules have been applied exhaustively and none of their conditions is applicable anymore. We first consider several branching rules that consider branching on vertices from $\Active$, not taking the special structure of chordal graphs into account.

\begin{table}[tbh]\centering 
\begin{tabular}{c|c|c}
   Rule  & branching vector & branching number\\ \hline
   \ref{3-in-\Active}       & $(1-\omega_2,1 + 3 \min(1-\omega_1,\omega_2))$                                            & 1.8940\\
   \ref{\Active-with-one-notV2-and-one-special-notV1} &$(1+\omega_1+\omega_2,1-\omega_2)$& 1.8014\\
   \ref{2-not-in-V1-bar}&$(\omega_1,\omega_1 + 2\omega_2)$&1.8940\\
   \ref{2-not-in-V1-bar-a}&$(\omega_1,2-\omega_1 + \min(1-\omega_1, \omega_2))$&1.8940\\
   \ref{simp-not-in-V1} \& \ref{simp-non-pendant-not-in-V2-3}&$(\omega_1, 2\omega_1+\omega_2)$&1.7915\\\ref{pendant-adjacent}
   &$(\omega_1+\omega_2,\omega_1+\omega_2)$&1.8321\\
   \ref{pendant-in-\Active-1} \& \ref{simp-non-pandant-not-in-V2-1}&$(1+\omega_2,1+\omega_2)$&\small never worse than \autoref{pendant-adjacent} \\
   \ref{pendant-in-\Active-1-a}&$(1+\omega_2+\min(1-\omega_2,\omega_1),1)$&1.6181\\
   \ref{pendant-in-\Active-2}&$(2,1-\omega_2)$&1.8471\\
   \ref{pendant-not-in-V2}&$(1+\omega_2,1)$&1.779\\
   \ref{pendant-not-in-V1-activ} & $(1+\omega_1+2(1-\omega_2), \omega_1)$& 1.5743\\
   \ref{simp-non-pendant-in-\Active-1} &$(1+2\omega_2,1)$&\small never worse than \autoref{pendant-not-in-V2}\\
   \ref{simp-non-pendant-in-\Active-2}&$(2+\omega_2,1-\omega_2)$&1.7249\\
   \ref{simp-non-pandant-not-in-V2-2}&$(2-\omega_1+\omega_2,2+\omega_2,2-\omega_2)$& 1.8005\\   
   \ref{semi-simp}&$(2-\omega_1+\omega_2,2+\omega_2,2-\omega_2)$& \small same as \autoref{simp-non-pandant-not-in-V2-2}\\
\end{tabular}
\caption{Branching rules and their vectors and numbers for chordal graphs}\label{tab:chordal-branchings}
\end{table}

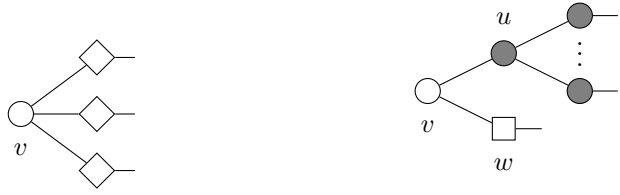
\begin{figure}[bt]
    \centering
    	
\begin{subfigure}[t]{.43\textwidth}
    \centering
    	
	\begin{tikzpicture}[transform shape]
			\tikzset{every node/.style={circle,minimum size=0.3cm}}
		\tikzset{every node/.style={ fill = white,circle,minimum size=0.3cm}}
			\node[draw,label={below:$v$}] (v) at (0,0) {};
			\node[draw,diamond] (u2) at (1,0.75) {};
			\node[draw,diamond] (w2) at (1,0) {};
			\node[draw,diamond] (u1) at (1,-0.75) {};
			\path (u1) edge[-] (v);
			\path (u2) edge[-] (v);
			\path (w2) edge[-] (v);
			\path (u1) edge[-] (1.5,-0.75);
			\path (w2) edge[-] (1.5,0);
			\path (u2) edge[-] (1.5,0.75);

        \end{tikzpicture}

    \subcaption{$v\in\Active$ has at least 3 neighbors in $\Active \cup \overline{V_2}$}
    \label{fig:3-in-\Active}
\end{subfigure}
\begin{subfigure}[t]{.51\textwidth}
    \centering
    	
	\begin{tikzpicture}[transform shape]
			\tikzset{every node/.style={ fill = white,circle,minimum size=0.3cm}}
			\node[draw,label={below:$v$}] (v) at (0,0) {};
			\node[draw,fill=gray,label={above:$u$}] (u2) at (1,0.5) {};
			\node[draw,rectangle,label={below:$w$}] (w2) at (1,-0.5) {};
			\node[draw,fill=gray] (u1) at (2,1) {};
			\node[draw,fill=gray] (u3) at (2,0) {};
			\node[fill=none] at (2,0.6) {$\vdots$};
			\path (u2) edge[-] (v);
			\path (u1) edge[-] (u2);
			\path (u2) edge[-] (u3);
			\path (w2) edge[-] (v);
			\path (w2) edge[-] (1.5,-0.5);
			\path (u1) edge[-] (2.5,1);
			\path (u3) edge[-] (2.5,0);

        \end{tikzpicture}

    \subcaption{$v\in \Active$ has one neighbor $w\in \overline{V_2}$ and at least one neighbor in $\overline{V_1}$ has only further neighbor in $\overline{V_1}$}
    \label{fig:\Active-with-one-notV2-and-one-special-notV1}
    \end{subfigure}

    \caption{Branching Rules~\ref{3-in-\Active} and~\ref{\Active-with-one-notV2-and-one-special-notV1}}
\end{figure}

\begin{brarule}\label{3-in-\Active}
If $v\in \Active$ has at least three neighbors in $\Active\cup \overline{V_2}$, then we branch as follows:

\begin{enumerate}
\item Set $f(v)=2$. Each neighbor $x\in N(v)\cap \Active$ is added to $\overline{V_1}$ and each element of $N(v)\cap \overline{V_2}$ is assigned a value of zero and deleted.

\item Add $v$ to $\overline{V_2}$.

\end{enumerate}
\end{brarule}

\begin{lemma}The case distinction of  \autoref{3-in-\Active} is complete. 
The worst-case branch vector is  $(1-\omega_2,1 + 3 \min(1-\omega_1,\omega_2))$.
\end{lemma}

\noindent
From now on, we can assume that a vertex from $\Active$ of degree three has a neighbor in~$\overline{V_1}$.

\begin{brarule}
\label{\Active-with-one-notV2-and-one-special-notV1}
If $v\in \Active$ has at least one neighbor~$w$ in $\overline{V_2}$ and at least one neighbor~$u$ in $\overline{V_1}$ such that all neighbors of $u$ (but $v$ and possibly $w$) are in $\overline{V_1}$, then we branch as follows:

\begin{enumerate}
    \item Set $f(v)=2$. Then, set all neighbors of $v$ that belong to $\overline{V_2}$ to zero (and delete them) and set all neighbors of $v$ that belong to $\overline{V_1}$ whose neighbors (but $v,w$) also belong to $\overline{V_1}$ also to zero (and delete them).
    \item Add $v$ to $\overline{V_2}$.
\end{enumerate}
\end{brarule}

\begin{lemma}
The case distinction of  \autoref{\Active-with-one-notV2-and-one-special-notV1} is complete. 
The worst-case branch vector is $(1 + \omega_1 +\omega_2, 1 - \omega_2)$.
\end{lemma}

\noindent
Knowing (by our invariants) that elements of $\overline{V_1}$ are guaranteed to have neighbors in $V_2$, the next two branching rules apply to some elements of $\overline{V_1}$:

\begin{brarule}\label{2-not-in-V1-bar}
If $v\in \overline{V_1}$ has at least two neighbors in $\overline{V_2}$, then we branch as follows:

\begin{enumerate}

\item Set $f(v)=2$. Each neighbor $x\in N(v)\cap \Active$ is added to $\overline{V_1}$ and each element of $N(v)\cap \overline{V_2}$ is assigned a value of zero and deleted.

\item Set $f(v)=0$; delete $v$.

\end{enumerate}
\end{brarule}

\begin{lemma}
The case distinction of  \autoref{2-not-in-V1-bar} is complete. 
The worst-case branch vector is $(\omega_1,\omega_1 + 2\omega_2)$.
\end{lemma}

\noindent

\begin{brarule}\label{2-not-in-V1-bar-a}
If $v\in \overline{V_1}$ has at least three neighbors in $\Active\cup \overline{V_2}$ then we branch as follows:

\begin{enumerate}

\item Set $f(v)=2$. Each neighbor $x\in N(v)\cap \Active$ is added to $\overline{V_1}$ and each element of $N(v)\cap \overline{V_2}$ is assigned a value of zero and deleted.

\item Set $f(v)=0$; delete $v$.

\end{enumerate}
\end{brarule}

\begin{lemma}The case distinction of  \autoref{2-not-in-V1-bar-a} is complete. 
The worst-case branch vector is $(\omega_1,\omega_1 + \min(1-\omega_1, \omega_2) + 2(1-\omega_1))$.
\end{lemma}

The worst case branch vector is given like this because $v$ has at least two neighbors in $\Active$ (otherwise \autoref{2-not-in-V1-bar} would be applied).

\begin{figure}[bt]
    \centering
    	
\begin{subfigure}[t]{.43\textwidth}
    \centering
    	
	\begin{tikzpicture}[transform shape]
			\tikzset{every node/.style={circle,minimum size=0.3cm}}
		\tikzset{every node/.style={ fill = white,circle,minimum size=0.3cm}}
			\node[draw,label={below:$v$},fill=gray] (v) at (0,0) {};
			\node[draw,rectangle] (u2) at (1,0.5) {};
			\node[draw,rectangle] (u1) at (1,-0.5) {};
			\node[fill=none] at (1,0.6) {};
			\path (u1) edge[-] (v);
			\path (u2) edge[-] (v);
			\path (u1) edge[-] (1.5,-0.5);
			\path (u2) edge[-] (1.5,0.5);

        \end{tikzpicture}

    \subcaption{$v\in \overline{V_1}$ has at least 2 neighbor $\overline{V_2}$}
    \label{fig:2-not-in-V1-bar}
\end{subfigure}
\begin{subfigure}[t]{.51\textwidth}
    \centering
    	
	\begin{tikzpicture}[transform shape]
			\tikzset{every node/.style={ fill = white,circle,minimum size=0.3cm}}
			\node[draw,fill=gray,label={below:$v$}] (v) at (0,0) {};
			\node[draw,diamond] (u2) at (1,0.75) {};
			\node[draw,diamond] (w2) at (1,0) {};
			\node[draw,diamond] (u1) at (1,-0.75) {};
			\path (u1) edge[-] (v);
			\path (u2) edge[-] (v);
			\path (w2) edge[-] (v);
			\path (u1) edge[-] (1.5,-0.75);
			\path (w2) edge[-] (1.5,0);
			\path (u2) edge[-] (1.5,0.75);

        \end{tikzpicture}

    \subcaption{$v\in \overline{V_1}$ has at least 3 neighbors in $\Active \cup \overline{V_2}$
    }
    \label{fig:2-not-in-V1-bar-a}
    \end{subfigure}

    \caption{Branching Rules~\ref{2-not-in-V1-bar} and~\ref{2-not-in-V1-bar-a}}
\end{figure}
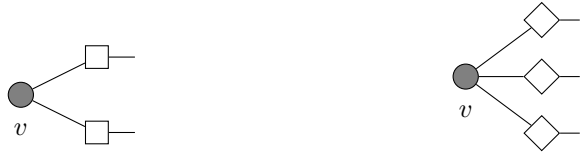
\noindent

From now on, we discuss branching on simplicial vertices (or sometimes of vertices in the neighborhood of simplicial vertices).

\begin{observation}
Notice that simplicial vertices in $\overline{V_1}$ can only have neighbors in $\Active\cup \overline{V_2}\cup\overline{V_1}$.
As we already considered the case of vertices in $\overline{V_1}$ that have at least three neighbors in $\Active\cup \overline{V_2}$, in the following branchings, we know that a vertex in~$\overline{V_1}$ has at most two neighbors in $\Active\cup \overline{V_2}$, not both of them being in $\overline{V_2}$ due to \autoref{2-not-in-V1-bar}.

\end{observation}

\begin{brarule}\label{simp-not-in-V1}
If $v\in \overline{V_1}$ is simplicial and of degree at least two, then we branch as follows:

\begin{enumerate}
    \item Set $f(v)=2$ and assigning a value of zero to each of its neighbors, which results in deleting $N[v]$.
    \item Set $f(v) =0$ and delete $v$.
\end{enumerate}

\end{brarule}

\begin{lemma}The case distinction of  \autoref{simp-not-in-V1} is complete. 
The worst-case branch vector is $(\omega_1, 2\omega_1+\omega_2)$.
\end{lemma}

\noindent
The branching on simplicial vertices is always better if some neighbors are in~$\Active$ (instead of being in $\overline{V_1}\cup\overline{V_2}$), because (due to $v$ being simplicial) a neighbor from $\Active$ would immediately move to $V_0$ in this branch. Hence, due to \autoref{2-not-in-V1-bar-a}, we can assume that $v$ has at least one neighbor in $\overline{V_1}$. This implies the worst case branching vector. It corresponds to the case where $v$ is of degree two  because of \autoref{2-not-in-V1-bar}. More specifically, $v$ has one neighbor in~$\overline{V_1}$ and one neighbor in $\overline{V_2}\cup \Active$ by \autoref{not-V2-with not-V2-neighbors}. Actually, in this situation, the worst case comes from a neighbor of $v$ from $\overline{V_2}$.

\noindent
Next we consider pendant simplicial vertices (and sometimes slightly more general situations). 

\begin{observation}
We first note that an isolated pair of pendant adjacent vertices, say $v,w$, give rise to a path, which has already been studied. However, assuming previous branching rules have resulted in such a path, the worst case here corresponds to $v\in \overline{V_2}$ and $w\in \overline{V_1}$. To see this, note that when both $v$ and $w$ are in~$\overline{V_2}$ or both in~$\overline{V_1}$, they would automatically be deleted by Reduction Rules \ref{not-V2-with not-V2-neighbors} or \ref{isolated-not-in-V1}. 

\end{observation}

\begin{brarule}\label{pendant-adjacent}
If $v\in \overline{V_2}$ is a vertex with exactly one neighbor $w\in \overline{V_1}$ and possibly more neighbors in $\overline{V_2}$, then we branch as follows:

\begin{enumerate}
    \item Set $f(w)=2$ and $f(v)=0$ (similarly, other neighbors of~$w$ are updated). 
    \item Set $f(w)=0$ and
    $f(v)=1$ (by \autoref{not-V2-with not-V2-neighbors}). 
\end{enumerate}
\end{brarule}

\begin{lemma}The case distinction of  \autoref{pendant-adjacent} is complete. 
The worst-case branch vector is $(\omega_1+\omega_2,\omega_1+\omega_2)$.
\end{lemma}

\noindent Notice that this implies $\omega_1+\omega_2>1$. Hence, \autoref{pendant-in-\Active-2} has a branching that is never better than that of \autoref{\Active-with-one-notV2-and-one-special-notV1}.

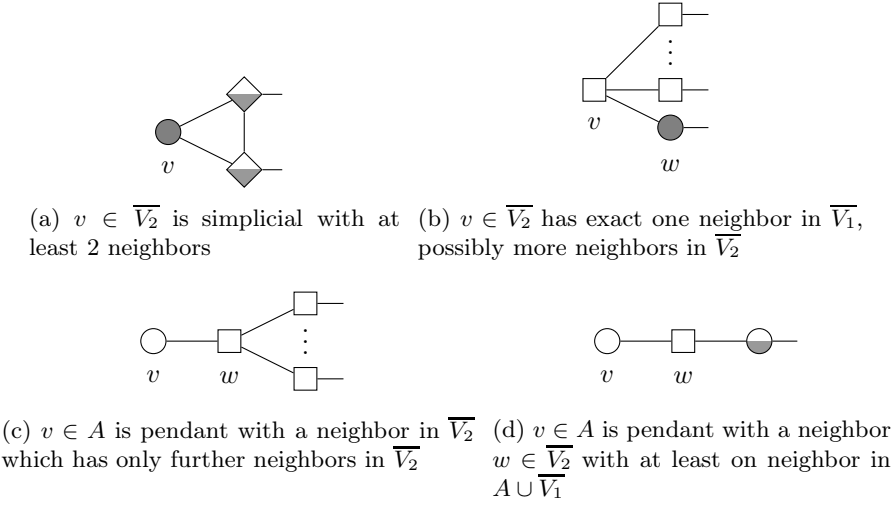
\begin{figure}[bt]
    \centering
    	
\begin{subfigure}[t]{.4\textwidth}
    \centering
    	
	\begin{tikzpicture}[transform shape]
			\tikzset{every node/.style={ fill = white,circle,minimum size=0.3cm},fill lower half/.style={path picture={\fill[#1] (path picture bounding box.south west) rectangle (path picture bounding box.east);}}}
		\tikzset{every node/.style={ fill = white,circle,minimum size=0.3cm}}
			\node[draw,label={below:$v$},fill=gray] (v) at (0,0) {};
			\node[draw,fill lower half=black!40,diamond] (u2) at (1,0.5) {};
			\node[draw,fill lower half=black!40,diamond] (u1) at (1,-0.5) {};
			\node[fill=none] at (1,0.6) {};
			\path (u1) edge[-] (v);
			\path (u2) edge[-] (v);
			\path (u1) edge[-] (u2);
			\path (u1) edge[-] (1.5,-0.5);
			\path (u2) edge[-] (1.5,0.5);

        \end{tikzpicture}

    \subcaption{$v\in \overline{V_2}$ is simplicial with at least 2 neighbors}
    \label{fig:simp-not-in-V1}
\end{subfigure}\ \ 
\begin{subfigure}[t]{.48\textwidth}
    \centering
    	
	\begin{tikzpicture}[transform shape]
			\tikzset{every node/.style={ fill = white,circle,minimum size=0.3cm}}
			\node[draw,label={below:$v$},rectangle] (v) at (0,0) {};
			\node[draw,rectangle,label={}] (u2) at (1,1) {};
			\node[draw,fill=gray,label={below:$w$}] (w2) at (1,-0.5) {};
			\node[draw,rectangle] (u1) at (1,0) {};
			\node[fill=none] at (1,0.6) {$\vdots$};
			\path (u2) edge[-] (v);
			\path (u1) edge[-] (v);
			\path (w2) edge[-] (v);
			\path (w2) edge[-] (1.5,-0.5);
			\path (u2) edge[-] (1.5,1);
			\path (u1) edge[-] (1.5,0);

        \end{tikzpicture}

    \subcaption{$v\in \overline{V_2}$ has exact one neighbor in $\overline{V_1}$, possibly more neighbors in $\overline{V_2}$}
    \label{fig:pendant-adjacent}
    \end{subfigure}
\begin{subfigure}[t]{.51\textwidth}
    \centering
    	
	\begin{tikzpicture}[transform shape]
			\tikzset{every node/.style={ fill = white,circle,minimum size=0.3cm}}
			\node[draw,label={below:$v$}] (v) at (0,0) {};
			\node[draw,rectangle,label={}] (u2) at (2,0.5) {};
			\node[draw,rectangle,] (w2) at (2,-0.5) {};
			\node[draw,rectangle,label={below:$w$}] (u1) at (1,0) {};
			\node[fill=none] at (2,0.1) {$\vdots$};
			\path (u1) edge[-] (v);
			\path (u1) edge[-] (u2);
			\path (w2) edge[-] (u1);
			\path (w2) edge[-] (2.5,-0.5);
			\path (u2) edge[-] (2.5,0.5);

        \end{tikzpicture}

    \subcaption{$v\in \Active$ is pendant with a neighbor in $\overline{V_2}$ which has only further neighbors in $\overline{V_2}$}
    \label{fig:pendant-in-\Active-1}
    \end{subfigure}\ \ 
    \begin{subfigure}[t]{.43\textwidth}
    \centering
    	
	\begin{tikzpicture}[transform shape]
			\tikzset{every node/.style={ fill = white,circle,minimum size=0.3cm},fill lower half/.style={path picture={\fill[#1] (path picture bounding box.south west) rectangle (path picture bounding box.east);}}}
		    \tikzset{every node/.style={ fill = white,circle,minimum size=0.3cm}}
			\node[draw,label={below:$v$}] (v) at (0,0) {}; {};
			\node[draw,fill lower half=black!40] (w2) at (2,0) {};
			\node[draw,rectangle,label={below:$w$}] (u1) at (1,0) {};
			\path (u1) edge[-] (v);
			\path (w2) edge[-] (u1);
			\path (w2) edge[-] (2.5,0);
        \end{tikzpicture}

    \subcaption{$v\in \Active$ is pendant with a neighbor $w\in \overline{V_2}$ with at least on neighbor in $\Active \cup \overline{V_1}$}
    \label{fig:pendant-in-\Active-1-a}
    \end{subfigure}
    \caption{Branching Rules~\ref{simp-not-in-V1}, \ref{pendant-adjacent}, \ref{pendant-in-\Active-1} and~\ref{pendant-in-\Active-1-a}}
\end{figure}

\begin{brarule}\label{pendant-in-\Active-1}
If $v\in \Active$ is of degree one and its neighbor $w\in \overline{V_2}$,
and assume that all neighbors of $w$ (but $v$) are also in $\overline{V_2}$, then we branch as follows:

\begin{enumerate}
    \item Set $f(v) = 2$ and $f(w) = 0$. 
    \item Set $f(v)=f(w)=1$ and delete them (implicitly by  \autoref{not-V2-with not-V2-neighbors}).
\end{enumerate}
\end{brarule}

\begin{lemma}The case distinction of  \autoref{pendant-in-\Active-1} is complete. 
The worst-case branch vector is $(1+\omega_2,1+\omega_2)$.
\end{lemma}

\noindent
This branching is never worse than that of \autoref{pendant-adjacent}.

\begin{brarule}\label{pendant-in-\Active-1-a}
If $v\in \Active$ is of degree one and its neighbor $w\in \overline{V_2}$,
and assume that there is at least one further neighbor of $w$  that belongs to $\Active\cup \overline{V_1}$,  then we branch as follows:

\begin{enumerate}
    \item Set $f(v) = 2$ and $f(w) = 0$ and, additionally, update all neighbors of $w$ to $\overline{V_2}$ or to $V_0$. 
    \item Set $f(v)=1$ and delete it (implicitly by  \autoref{not-V2-with not-V2-neighbors}).
\end{enumerate}
\end{brarule}

\begin{lemma}
The case distinction of  \autoref{pendant-in-\Active-1-a} is complete. 
The worst-case branch vector is $(1+\omega_2+\min(1-\omega_2,\omega_1),1)$.
\end{lemma}

\noindent
The reason for this branching vector is that one such neighbor must exist due to the previous branching rule.

The following rule again deals with a pendant vertex as a special case, but we allow some more vertices to come into the play, as it helps solve another situation turning up later.

\begin{figure}[bt]
    \centering
    	
\begin{subfigure}[t]{.22\textwidth}
    \centering
    	
	\begin{tikzpicture}[transform shape]
			\tikzset{every node/.style={ fill = white,circle,minimum size=0.3cm},fill lower half/.style={path picture={\fill[#1] (path picture bounding box.south west) rectangle (path picture bounding box.east);}}}
		\tikzset{every node/.style={ fill = white,circle,minimum size=0.3cm}}
			\node[draw,label={below:$v$}] (v) at (0,0) {};
			\node[draw,fill=gray] (u2) at (1,1) {};
			\node[draw,label={below:$w$}] (w2) at (1,-0.5) {};
			\node[draw,fill=gray] (u1) at (1,0) {};
			\node[fill=none] at (1,0.6) {$\vdots$};
			\path (u2) edge[-] (v);
			\path (u1) edge[-] (v);
			\path (w2) edge[-] (v);
			\path (w2) edge[-] (1.5,-0.5);
			\path (u2) edge[-] (1.5,1);
			\path (u1) edge[-] (1.5,0);

        \end{tikzpicture}

    \subcaption{$v\in \Active$ with exactly one neighbor in $\Active$ and possible other neighbor in~$\overline{V_1}$}
    \label{fig:pendant-in-\Active-2}
\end{subfigure}\ \ 
\begin{subfigure}[t]{.22\textwidth}
    \centering
    	
	\begin{tikzpicture}[transform shape]
			\tikzset{every node/.style={ fill = white,circle,minimum size=0.3cm}}
			\node[draw,label={below:$v$},rectangle] (v) at (0,0) {};
			\node[draw] (u1) at (1,0) {};
			\path (u1) edge[-] (v);
			\path (u1) edge[-] (1.5,0);

        \end{tikzpicture}

    \subcaption{$v\in \overline{V_2}$ is pendant with a neighbor in $\Active$}
    \label{fig:pendant-not-in-V2}
    \end{subfigure}\ \ 
\begin{subfigure}[t]{.22\textwidth}
    \centering
    	
	\begin{tikzpicture}[transform shape]
			\tikzset{every node/.style={ fill = white,circle,minimum size=0.3cm}}
			\node[draw,label={below:$v$}] (v) at (0,0) {};
			\node[draw,rectangle,label={}] (u2) at (1,0.75) {};
			\node[draw,rectangle,] (w2) at (1,-0.75) {};
			\node[fill=none] at (0.7,0.1) {$\vdots$};
			\path (v) edge[-] (u2);
			\path (w2) edge[-] (v);
			\path (w2) edge[-] (u2);
			\path (w2) edge[-] (1.5,-0.75);
			\path (u2) edge[-] (1.5,0.75);

        \end{tikzpicture}

    \subcaption{$v\in \Active$ is simplicial and has only neighbors in $\overline{V_2}$}
    \label{fig:simp-non-pendant-in-\Active-1}
    \end{subfigure}\ \ 
    \begin{subfigure}[t]{.22\textwidth}
    \centering
    	
	\begin{tikzpicture}[transform shape]
			\tikzset{every node/.style={ fill = white,circle,minimum size=0.3cm},fill lower half/.style={path picture={\fill[#1] (path picture bounding box.south west) rectangle (path picture bounding box.east);}}}
		    \tikzset{every node/.style={ fill = white,circle,minimum size=0.3cm}}
            \node[draw,label={below:$v$}] (v) at (0,0) {};
			\node[draw,fill lower half=gray, diamond] (u2) at (1,1) {};
			\node[draw,label={below:$w$}] (w2) at (1,-0.5) {};
			\node[draw,fill lower half=gray, diamond] (u1) at (1.7,0) {};
			\node[fill=none] at (1.2,0.5) {$\vdots$};
			\path (u2) edge[-] (v);
			\path (u1) edge[-] (v);
			\path (w2) edge[-] (v);
			\path (u2) edge[-] (w2);
			\path (u1) edge[-] (u2);
			\path (w2) edge[-] (u1);
			\path (w2) edge[-] (1.5,-0.5);
			\path (u2) edge[-] (1.5,1);
			\path (u1) edge[-] (2.2,0);

        \end{tikzpicture}

    \subcaption{$v\in \Active$ is a simplicial vertex of degree at least 3 and at least one neighbor in $\Active$}
    \label{fig:simp-non-pendant-in-\Active-2}
    \end{subfigure}
    \caption{Branching Rules~\ref{pendant-in-\Active-2}, \ref{pendant-not-in-V2}, \ref{simp-non-pendant-in-\Active-1} and~\ref{simp-non-pendant-in-\Active-2}}
\end{figure}
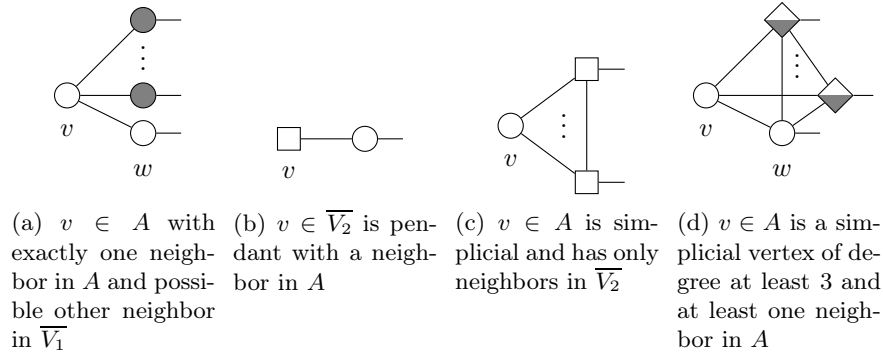

\begin{brarule}\label{pendant-in-\Active-2}
If $v\in \Active$ has one neighbor $w\in \Active$ and possibly other neighbors in $\overline{V_1}$, then we branch as follows:

\begin{enumerate}
    \item Set $f(v) = 2$ and $f(w) = 0$ (and put all further neighbors of $w$ into $\overline{V_2}$ or even into $V_0$; then, we can delete $v,w$).
    \item Add $v$ to $\overline{V_2}$.
\end{enumerate}
\end{brarule}

\begin{lemma}The case distinction of  \autoref{pendant-in-\Active-2} is complete. 
The worst-case branch vector is $(2,1-\omega_2)$.
\end{lemma}
\begin{pf}
The correctness of this rule follows, because when we set $f(v)=2$, then vertex~$v$ needs a private neighbor, and since all neighbors of $v$ (if any) but~$w$ belong to~$\overline{V_1}$ by assumption, $w$ must be the private neighbor of~$v$, which also means that no other neighbor of $w$ can be set to two.
\end{pf}

\begin{brarule}\label{pendant-not-in-V2}
If $v\in \overline{V_2}$ is of degree one and its neighbor $w\in \Active$, then we branch as follows:

\begin{enumerate}
    \item Set $f(w) = 2$ and $f(v) = 0$ (and add all other neighbors of $w$ to $\overline{V_1}$ or to~$V_0$).
    \item Add $w$ to $\overline{V_2}$, set $f(v)=1$ and delete $v$.
\end{enumerate}
\end{brarule}

\begin{brarule}\label{pendant-not-in-V1-activ}
If $v\in \overline{V_1}$ is of degree one and its neighbor $w\in \Active$ with 2 neighbors in $\Active$, then we branch as follows:

\begin{enumerate}
    \item Set $f(v) = 2$ and $f(w) = 0$ and put the neighbors of $w$ into $\overline{V_2}$.
    \item Set $f(v)=0$ and delete $v$.
\end{enumerate}
\end{brarule}

\begin{lemma}
The case distinction of  \autoref{pendant-not-in-V1-activ} is complete. 
The worst-case branch vector is $(1+\omega_1+2(1-\omega_2), \omega_1)$.
\end{lemma} 
\noindent

\begin{lemma}
The case distinction of  \autoref{pendant-not-in-V2} is complete. 
The worst-case branch vector is $(1+\omega_2,1)$.
\end{lemma} 
\noindent
From this point on, every simplicial vertex of degree more than one must be in $\Active\cup \overline{V_2}$. Moreover, a simplicial vertex from $\Active$ cannot have a neighbor in $\overline{V_1}$, otherwise either \autoref{all-neighbors-not-in-V1} or \autoref{2-not-in-V1-bar} would have been applied.

\begin{brarule}\label{simp-non-pendant-in-\Active-1}
If $v\in \Active$ is simplicial, of degree at least two such that $N(v)\subseteq \overline{V_2}$, then we branch as follows:

\begin{enumerate}
    \item Set $f(v) = 2$ and assign zero to all its neighbors (delete $N[v]$).
    \item Set $f(v)=1$ and delete it.
    \end{enumerate}
\end{brarule}

\begin{lemma}The case distinction of \autoref{simp-non-pendant-in-\Active-1} is complete. 
The worst-case branch vector is $(1+2\omega_2,1)$.
\end{lemma} 

\noindent
This branch is never worse than \autoref{pendant-not-in-V2}. 

\begin{brarule}\label{simp-non-pendant-in-\Active-2}
If $v\in \Active$ is simplicial, of degree at least two, with a neighbor $w\in \Active$, then we branch as follows:

\begin{enumerate}
    \item Set $f(v) = 2$ and assign zero to all its neighbors (delete $N[v]$).
    \item Add $v$ to $\overline{V_2}$.    \end{enumerate}
\end{brarule}

\begin{lemma}
The case distinction of \autoref{simp-non-pendant-in-\Active-2} is complete. 
The worst-case branch vector is $(2+\omega_2,1-\omega_2)$.
\end{lemma}

\noindent
Finally, we consider simplicial vertices in $\overline{V_2}$ of degree two or more. Many cases are already dealt with. The following rules cover the remaining cases.

\begin{observation}

Let $v\in \overline{V_2}$ be a simplicial vertex of degree two or more. Then $v$ must have at least one neighbor $w\in \Active\cup \overline{V_1}$ due to \autoref{not-V2-with not-V2-neighbors}. If $w\in \Active$, then at most one other neighbor of $v$ can be in $\Active\cup \overline{V_2}$ (by \autoref{3-in-\Active}) and no other neighbor of $v$ can be in $\overline{V_1}$ (by \autoref{\Active-with-one-notV2-and-one-special-notV1}). In this case, the degree of $v$ must be exactly two.
On the other hand, if $w\in \overline{V_1}$, then (evading the previous case) $N(v)\subseteq \overline{V_1}$, because if $v$ would have any neighbors in $\overline{V_2}$, then \autoref{pendant-adjacent} would apply.

\end{observation}

\begin{brarule}\label{simp-non-pandant-not-in-V2-1}
Let $v\in \overline{V_2}$ be a simplicial vertex of degree two with a neighbor $w\in \Active$. 
If the other neighbor $w'$ of $v$ is in $\overline{V_2}$, then we branch as follows:

\begin{enumerate}
    
    \item Set $f(w) = 2$ and $f(v)= f(w') = 0$. 
    
    \item Add $w$ to $\overline{V_2}$, set $f(v) = 1$ and finally delete it according to \autoref{not-V2-with not-V2-neighbors}.
    \end{enumerate}
\end{brarule}

\begin{lemma}The case distinction of \autoref{simp-non-pandant-not-in-V2-1} is complete. 
The worst-case branch vector is $(1+2\omega_2,1)$.
\end{lemma}

\noindent
\autoref{simp-non-pandant-not-in-V2-1} has the same branching vector as \autoref{simp-non-pendant-in-\Active-1}.

\begin{brarule}\label{simp-non-pandant-not-in-V2-2}
If $v\in \overline{V_2}$ be simplicial with two neighbors $w,w'\in \Active$, then we branch as follows:

\begin{enumerate}
    \item Set $f(w) = 2$, $f(v) = 0$ and add $w'$ to $\overline{V_1}$.
    \item Set $f(w')=2$ and $f(w)=f(v)=0$.
    \item Add $w$ and $w'$ to $\overline{V_2}$ and set $f(v)=1$.
    \end{enumerate}
\end{brarule}

\begin{lemma}The case distinction of \autoref{simp-non-pandant-not-in-V2-2} is complete. 
The worst-case branch vector is $(2-\omega_1+\omega_2,2+\omega_2,2-\omega_2)$.
\end{lemma}

\noindent 
This branching also works if $v,w,w'$ form a connected (triangle) component.

\begin{observation}
We now discuss a simplicial vertex $v\in \overline{V_2}$ of degree at least two with a neighbor $w\in \overline{V_1}$.
If $w$ has any other neighbor $\Active\in \overline{V_2}$, then \autoref{2-not-in-V1-bar} would trigger.
If $w$ has any neighbor $u\in \Active$, then  $u$ must have at least one neighbor~$u'$ that is not in $\overline{V_1}$ (by \autoref{all-neighbors-not-in-V1}). If $u'\in \overline{V_2}$, then \autoref{\Active-with-one-notV2-and-one-special-notV1} would be applicable. Hence, $u'\in \Active$, but this case is resolved by \autoref{pendant-in-\Active-2}. Hence, all neighbors of $w$ (but $v$) must belong to~$\overline{V_1}$.
\end{observation}

\begin{figure}[bt]
    \centering
    	
\begin{subfigure}[t]{.24\textwidth}
    \centering
    	
	\begin{tikzpicture}[transform shape]
			\tikzset{every node/.style={ fill = white,circle,minimum size=0.3cm},fill lower half/.style={path picture={\fill[#1] (path picture bounding box.south west) rectangle (path picture bounding box.east);}}}
		\tikzset{every node/.style={ fill = white,circle,minimum size=0.3cm}}
			\node[draw,label={below:$v$},rectangle] (v) at (0,0) {};
			\node[draw,label={above:$w$}] (u2) at (1,0.75) {};
			\node[draw,rectangle,] (w2) at (1,-0.75) {};
			\path (v) edge[-] (u2);
			\path (w2) edge[-] (v);
			\path (w2) edge[-] (u2);
			\path (w2) edge[-] (1.5,-0.75);
			\path (u2) edge[-] (1.5,0.75);

        \end{tikzpicture}

    \subcaption{$v\in \overline{V_2}$ is a simplicial vertex with exactly 1 neighbor in $\Active$ and 1 in $\overline{V_2}$}
    \label{fig:simp-non-pandant-not-in-V2-1}
\end{subfigure}\ \ 
\begin{subfigure}[t]{.24\textwidth}
    \centering
    	
	\begin{tikzpicture}[transform shape]
			\tikzset{every node/.style={ fill = white,circle,minimum size=0.3cm}}
			\node[draw,label={below:$v$},rectangle] (v) at (0,0) {};
			\node[draw,label={above:$w$}] (u2) at (1,0.75) {};
			\node[draw,label={below:$w'$}] (w2) at (1,-0.75) {};
			\path (v) edge[-] (u2);
			\path (w2) edge[-] (v);
			\path (w2) edge[-] (u2);
			\path (w2) edge[-] (1.5,-0.75);
			\path (u2) edge[-] (1.5,0.75);

        \end{tikzpicture}

    \subcaption{$v\in \overline{V_2}$ is a simplicial vertex with only 2 neighbors in $\Active$ }
    \label{fig:simp-non-pandant-not-in-V2-2}
    \end{subfigure}\ \ 
\begin{subfigure}[t]{.42\textwidth}
    \centering
    	
	\begin{tikzpicture}[transform shape]
			\tikzset{every node/.style={ fill = white,circle,minimum size=0.3cm},fill lower half/.style={path picture={\fill[#1] (path picture bounding box.south west) rectangle (path picture bounding box.east);}}}
			\node[draw,rectangle,label={below:$v$}] (v) at (0,0) {};
			\node[draw,fill=gray,label={above:$w$}] (u2) at (1,0.5) {};
			\node[draw,diamond,fill lower half=gray] (w2) at (1,-0.5) {};
			\node[draw,fill=gray] (u1) at (2,1) {};
			\node[draw,fill=gray] (u3) at (2,0) {};
			\node[fill=none] at (2,0.6) {$\vdots$};
			\path (u2) edge[-] (v);
			\path (u2) edge[-] (w2);
			\path (u1) edge[-] (u2);
			\path (u2) edge[-] (u3);
			\path (w2) edge[-] (v);
			\path (w2) edge[-] (1.5,-0.5);
			\path (u1) edge[-] (2.5,1);
			\path (u3) edge[-] (2.5,0);

        \end{tikzpicture}

    \subcaption{$v\in \overline{V_2}$ is a simplicial vertex  of degree two with 1 neighbor $w\in \overline{V_1}$ such that $N[w]\setminus \{ v\}\subseteq \overline{V_1}$ }
    \label{fig:simp-non-pendant-not-in-V2-3}
    \end{subfigure}
    \caption{Branching Rules~\ref{simp-non-pandant-not-in-V2-1}, \ref{simp-non-pandant-not-in-V2-2} and~\ref{simp-non-pendant-not-in-V2-3}}
\end{figure}
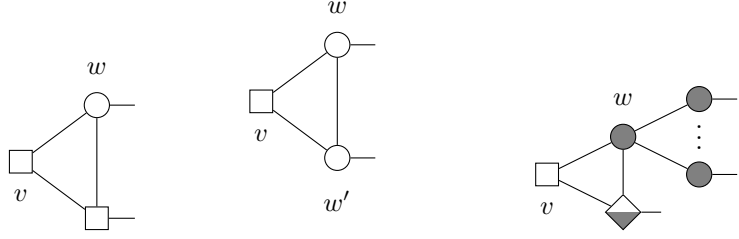

\begin{brarule}\label{simp-non-pendant-not-in-V2-3}
If $v\in \overline{V_2}$ is simplicial, of degree at least two, with a neighbor~$w$ such that $N[w]\setminus\{v\}\subseteq \overline{V_1}$, then we branch as follows:

\begin{enumerate}
    \item Set $f(w) = 2$, $f(v) = 0$ and delete $N[v]$.
    \item Set $f(w)=0$ and delete it.
    \end{enumerate}
\end{brarule}

\begin{lemma}
The case distinction of \autoref{simp-non-pendant-not-in-V2-3} is complete. 
The worst-case branch vector is $(2\omega_1+\omega_2,\omega_1)$.
\end{lemma}
\noindent
In the first branch, observe that by our previous reasoning, neighbors of $v$ are in $\overline{V_1}$, and do not have any more vertices to dominate. This is the same as in \autoref{simp-not-in-V1}.

\begin{brarule}\label{semi-simp}
If $v\in \overline{V_2}$ is simplicial, of degree at least two, with neighbors $w,w'\in \overline{V_1}$ such that $N[w]\subseteq N[w']$, then we branch as follows:

\begin{enumerate}
    \item Set $f(w') = 2$, $f(v)=f(w) = 0$ .
    \item Set $f(w')=0$ and delete it.
    \end{enumerate}
\end{brarule}

\begin{lemma}
The case distinction of \autoref{semi-simp} is complete. 
The worst-case branch vector is $(2\omega_1+\omega_2,\omega_1)$.
\end{lemma}

\begin{lemma}
The Reduction and Branching Rules cover all possible cases for chordal graphs.
\end{lemma}

\begin{pf}
Since each chordal graph has at least one simplicial vertex, we will show that each case for simplical vertex $v\in \Active\cup \overline{V_1}\cup \overline{V_2}$ is covered. At first let $v\in \Active$. By the branching rules \ref{pendant-in-\Active-1}, \ref{pendant-in-\Active-1-a} and \ref{pendant-in-\Active-2}, $v$ can not be pendant. $N(v)\subseteq \overline{V_2}$ triggers \autoref{simp-non-pendant-in-\Active-1}. If $N(v)\cap \Active$ is not empty it would trigger \ref{simp-non-pendant-in-\Active-2}. The remaining cases are covered by \autoref{\Active-with-one-notV2-and-one-special-notV1}.

A simplicial vertex $v\in \overline{V_1}$ with degree at least 2 would trigger~\autoref{simp-not-in-V1}. A pendant $v\in \overline{V_1}$  has to have a neighbor in $w\in \Active \cup \overline{V_2}$. $N[w]\setminus\{v\}\subseteq \overline{V_2}$ triggers \autoref{pendant-adjacent}. For $w\in \Active$ with $N[w]\cap \overline{V_2}\neq \emptyset$, we can use \autoref{\Active-with-one-notV2-and-one-special-notV1}. Because of \autoref{isolated-not-in-V1} and \autoref{pendant-in-\Active-2}, $w$ needs at least~2 neighbors in $\Active$. \autoref{3-in-\Active} prevents that $w$ has~3 neighbors in $\Active$. This triggers \autoref{pendant-not-in-V1-activ}. 

Let $v\in \overline{V_2}$. If $v$ is pendant, it triggers a reduction rule, \autoref{pendant-adjacent} or \autoref{pendant-not-in-V2}. If $v$ has degree~$2$ and one neighbor is in $\Active$, then this triggers one of the Branching Rules \ref{2-not-in-V1-bar}, ~\ref{simp-non-pandant-not-in-V2-1}, \ref{simp-non-pandant-not-in-V2-2} and~\ref{simp-non-pendant-not-in-V2-3}. 
Finally, assuming we have used all branching rules exhaustively, and particularly because of \autoref{\Active-with-one-notV2-and-one-special-notV1}, \autoref{2-not-in-V1-bar} and \autoref{2-not-in-V1-bar-a}, no two simplicial vertices in $\overline{V_2}$ can have a common neighbor. Therefore, at least one such simplicial vertex $v$ must have a semi-simplicial neighbor $w$. In other words, $w$ becomes simplicial after its deletion, as defined in \cite{AbuHeg2016}. This means $N[w]\subset N[w']$ for each other neighbor $w'$ of $v$, which justifies the branching in \autoref{semi-simp}.
\end{pf}  

\section{Concluding Remarks and Suggestions for Future Research}

The number of minimal Roman dominating functions in a graph of order $n$ has been recently shown, constructively, to be in $\mathcal{O}(\RomanUpperbound^n)$ \cite{AbuFerMan2022}. In this paper, we consider the same enumeration problem on special graph classes, namely split graphs, cobipartite graphs, forests, internal graphs, and (general) chordal graphs. 

On chordal graphs, we present an enumeration algorithm that runs in time $\mathcal{O}(\RomanChordalUpperbound^n)$, which gives a better upper bound on the number of minimal rdfs than in the general case.
We observe that the maximum number of minimal rdfs in forests and interval graphs (and more generally chordal graphs) of order $n$ is in 
$\Omega(\RomanIntervalExact^n)$. 
The lower bound in the case of split graphs and cobipartite graphs is shown to be in 
$\Omega(\sqrt{2}^n)$. 
We show that minimal rdfs can be enumerated in 
$\Omega(\RomanIntervalUpperbound^n)$ on forests and interval graphs, thus achieving an optimal input-sensitive enumeration in these cases. For split and cobipartite graphs our enumeration is not tight, but gives an upper bound of $\mathcal{O}(\RomanSplitUpperbound^n)$, which is relatively close to the lower bound. 

For future work, it would be an interesting natural question to close the gap between the presented lower and upper bounds in the case of split and cobipartite graphs. Our enumeration algorithm in the case of forests was more exhaustive and more intricate than one would expect for this graph class, yet it proved to be optimal as we noted above. It would be interesting to obtain a simplified enumeration algorithm in this case. 

More specifically, it remains open whether enumeration on chordal graphs can be improved further, so we hereby pose it as an open problem, or whether one can obtain a higher lower bound, which would also be a gap-improvement on general graphs.
So far, the best lower bound for general graphs is a collection of $C_5$, which is clearly not a chordal graph. The worst-case example for chordal graphs is a collection of $P_2$, see \autoref{sec:RomanPaths}.
Moreover, we did not see any way to improve the enumeration for bipartite graphs over the general case, which is a natural class to consider, given our results on split and on cobipartite graphs.

Generally speaking, 
there is also the combinatorial question of determining the number of minimal or maximal solutions for certain graph classes. While this question is generally \#\textsf{P}-hard, also on restricted graph classes (see~\cite{KanUno2017}), concrete recursive formulas are known for special graph classes as paths and cycles in the case of domination; see \cite[Propos. 7]{Bro2011} for the number of minimal dominating sets on a path, or \cite{Fur87} for the number of maximal independent sets on paths and cycles. More seems to be known about maximal independent sets compared to minimal dominating sets; we also refer to \cite{ChaJou99,GriGriGui88,KohGohDon2008}. In this paper, we developed a recursive formula for the number of minimal rdf on paths, but we did not look into general counting problems with respect to minimal rdf, nor did we look at other seemingly simple graph classes. Also in this respect, many tasks remain to be executed.

Many variants of Roman domination have been proposed in the literature, for instance, regarding
multi-attack variants, see \cite{Cocetal2005,GodHedHed2005,Hen2003,HenHed2003},
multiple Roman domination \cite{AbdCheShe2017,AbdACSV2021,BeeHayHed2016,MojVol2020} or so-called Italian domination and its variants, see \cite{CheHHM2016,DetLemRod2021,MojVol2020}. For all these variations, one could discuss the question of enumerating minimal dominating functions. This describes a vast open area of research. In particular, it would be interesting to see other problems related to Roman domination with a polynomial-time solvable extension variant.

\end{document}